\documentclass[fleqn,usenatbib]{mnras}

\usepackage{newtxtext,newtxmath}
\usepackage[T1]{fontenc}

\DeclareRobustCommand{\VAN}[3]{#2}
\let\VANthebibliography\thebibliography
\def\thebibliography{\DeclareRobustCommand{\VAN}[3]{##3}\VANthebibliography}

\usepackage{graphicx}	% Including figure files
\usepackage{subcaption} % Subfigures
\usepackage{amsmath}	% Advanced maths commands
\usepackage{siunitx}    % Easy SI units input
\usepackage{mathtools}  % Maths tools: \drcase{}
\usepackage{multirow}

\title[Orbital stability in the late stages of planetary evolution]{A Bayesian Monte Carlo assessment of orbital stability in the late stages of planetary system formation}

\author[J. Salas et al.]{
Jassyr Salas,$^{1}$\thanks{E-mail: sjassyr@mail.uniatlantico.edu.co}
Frank Bautista,$^{2}$
and Germ\'an Chaparro$^{3}$
\\
$^{1}$Universidad del Atl\'antico, Barranquilla, Colombia\\
$^{2}$Universidad Nacional de Colombia, Bogot\'a, Colombia\\
$^{3}$FACom,  
Instituto de F\'{\i}sica - FCEN, Universidad de Antioquia,
Calle 70 No. 52-21, Medell\'in, Colombia.\\
}

\date{Accepted XXX. Received YYY; in original form ZZZ}

\pubyear{2015}

\begin{document}
\label{firstpage}
\pagerange{\pageref{firstpage}--\pageref{lastpage}}
\maketitle

\begin{abstract}
The final orbital configuration of a planetary system is shaped by both its early star-disk environment and late-stage gravitational interactions. Assessing the relative importance of each of these factors is not straightforward due to the observed diversity of planetary systems compounded by observational biases. Our goal is to understand how a planetary system may change when planetesimal accretion and planet migrations stop and secular gravitational effects take over. Our approach starts with a novel classification of planetary systems based on their orbital architecture, validated using Approximate Bayesian Computation methods. We apply this scheme to observed planetary systems and also to $\sim400$ synthetic systems hosting $\sim5000$ planets, synthesized from a Monte Carlo planet population model. Our classification scheme robustly yields four system classes according to their planet masses and semi-major axes, for both observed and synthetic systems. We then estimate the orbital distribution density of each of the synthetic systems before and after dynamically evolving for 0.1-1 Myr using a gravitational+collisional $N$-body code. Using the Kullback-Leibler divergence to statistically measure orbital configuration changes, we find that $\lesssim$10\% of synthetic planetary systems experience such changes. We also find that this fraction belongs to a class of systems for which their center of mass is very close to their host star. Although changes in the orbital configuration of planetary systems may not be very common, they are more likely to happen in systems with close-in, massive planets, with F- and G-type host-stars and stellar metallicities [Fe/H] $>0.2$.

%This is a simple template for authors to write new MNRAS papers.
%The abstract should briefly describe the aims, methods, and main results of the paper.
%It should be a single paragraph not more than 250 words (200 words for Letters).
%No references should appear in the abstract.
\end{abstract}


%%%%%%%%%%%%%%%%%%%%%%%%%%%%%%%%%%%%%%%%%%%%%%%%%%

%%%%%%%%%%%%%%%%% BODY OF PAPER %%%%%%%%%%%%%%%%%%

\section{Introduction}\label{introduction}

Exoplanet catalogs have grown by significant leaps as new instruments have come operational \citep{bashi18,tess_2021} and the boundaries of the parameter space that defines exoplanetary system configurations has been constantly redefined ever since exoplanets were first detected \citep{udry07}. On the other hand, conclusions based on theories and numerical simulations developed for the early stages and the formation of the Solar System, explain how the dissipation of gas leads to instabilities that result in migrations and the formation of rocky planets \cite{Liu2022}, however, they are often difficult to apply to the formation history of other planetary systems \citep{RAYMOND2009644, WOO2022114692}. Additionally, exoplanet observations have biases that favor the detection of short-period planets. Therefore, deducing the physical conditions in which planetary systems are born based only on current observations is not straightforward.

However, despite the difficulty in overcoming detection biases, in recent years the number of detected exoplanetary systems appears to represent a partial but robust sampling of the parameter space for the physical conditions in which these systems form \citep{bryson_occur,bashi_2022}. Modern statistical methods can then be used to compare synthetic and observational results in order to make a meaningful taxonomy of planetary systems \citep{chaparro_2018,rogers_dist_planets}.

Deterministic models of the formation of planetary systems yield planetary system configurations that are highly sensitive to context-dependent effects such as dynamical instabilities and migration \citep{Ida_2004,schlecker_pps}. Therefore it is important to elucidate whether secular gravitational interactions play a role in the final configuration of a planetary system.

Here we present a study of planet formation in its final stages, when planetesimal accretion is mostly finished and protoplanets become gravitationally isolated planets \cite{beauge1990n,mordasini09_I}. This last stage is ruled by gravitational interactions between planets and planetesimals that tend to perturb their orbits when close encounters and collisions occur, until a final configuration of planets in differentiated stable orbits is achieved \citep{andryushin21}. This analysis can be done through computer simulations of synthetic planetary populations. Understanding this last phase is also of particular interest because the last impact record \cite{ALEXANDER1998113} and the possible formation of satellites are thought to emerge here \citep{alvarado_exomoons,sucerquia_exomoons}.

Therefore our main objective is to analyze the effect that these gravitational interactions have on the development and evolution of exoplanet populations, using a model of collisions between newly-accreted planets that depend on the impact energy between them. We propose a statistical framework to compare the initial and final configurations with a reconstruction of the posterior probability, which could help guide future exoplanet surveys.

We also implement a novel classification scheme for observational and synthetic planetary systems, which improves upon previous machine learning-based algorithms \cite{naderi19} in terms of adaptability and the fact that we focus on system-wide orbital characteristics. This is also of great help in understanding how the systems change after considering the gravitational interactions and evolution of exoplanets.

This work is structured as follows: firstly, in section \ref{data_treat} we go through the observational and synthetic data and describe our classification scheme. Then in Section \ref{methods} we describe the collisional $N$-body model used in the simulations and the statistical methods used for assessing system-wide orbital changes. In Section \ref{results} we discuss our results of the simulations, contrasting with the \cite{exoplanet} database, and finish with our conclusions in Section \ref{conclusion}.

% This is a simple template for authors to write new MNRAS papers.
% See \texttt{mnras\_sample.tex} for a more complex example, and \texttt{mnras\_guide.tex}
% for a full user guide.

% All papers should start with an Introduction section, which sets the work
% in context, cites relevant earlier studies in the field by %\citet{Fournier1901},
% and describes the problem the authors aim to solve %\citep[e.g.][]{vanDijk1902}.
% Multiple citations can be joined in a simple way like %\citet{deLaguarde1903, delaGuarde1904}.

\section{Planetary system data}\label{data_treat}
Our main focus in this work is to understand how the final orbital configuration of exoplanetary systems depends on stellar parameters and/or on secular gravitational effects. However, we first need to understand why observed planetary systems populations may differ from planetary population synthesis results. To this end, we use Unsupervised Learning and Approximate Bayesian Computation methods in order to compare observational and synthetic data, thus getting a grasp on the possible impact of observational biases. 
\subsection{Observational data}\label{obs_data}
Observations of exoplanetary systems give us a wide variety of data to benchmark planet formation models against. One of the most complete and public exoplanet databases is \cite{exoplanet} \citep{schneider11}, founded by the \textit{Observatoire de Paris}. The catalogue relates the physical parameters of exoplanets and the physical information of the host stars. In this database, sub\-stellar bodies with masses between $1\times10^{-5}$ (M$_\text{Jup}$) to $8\times10^{1}$ (M$_\text{Jup}$) are taken to be planets. 

\begin{figure}\label{}
	\includegraphics[width=0.9\columnwidth]{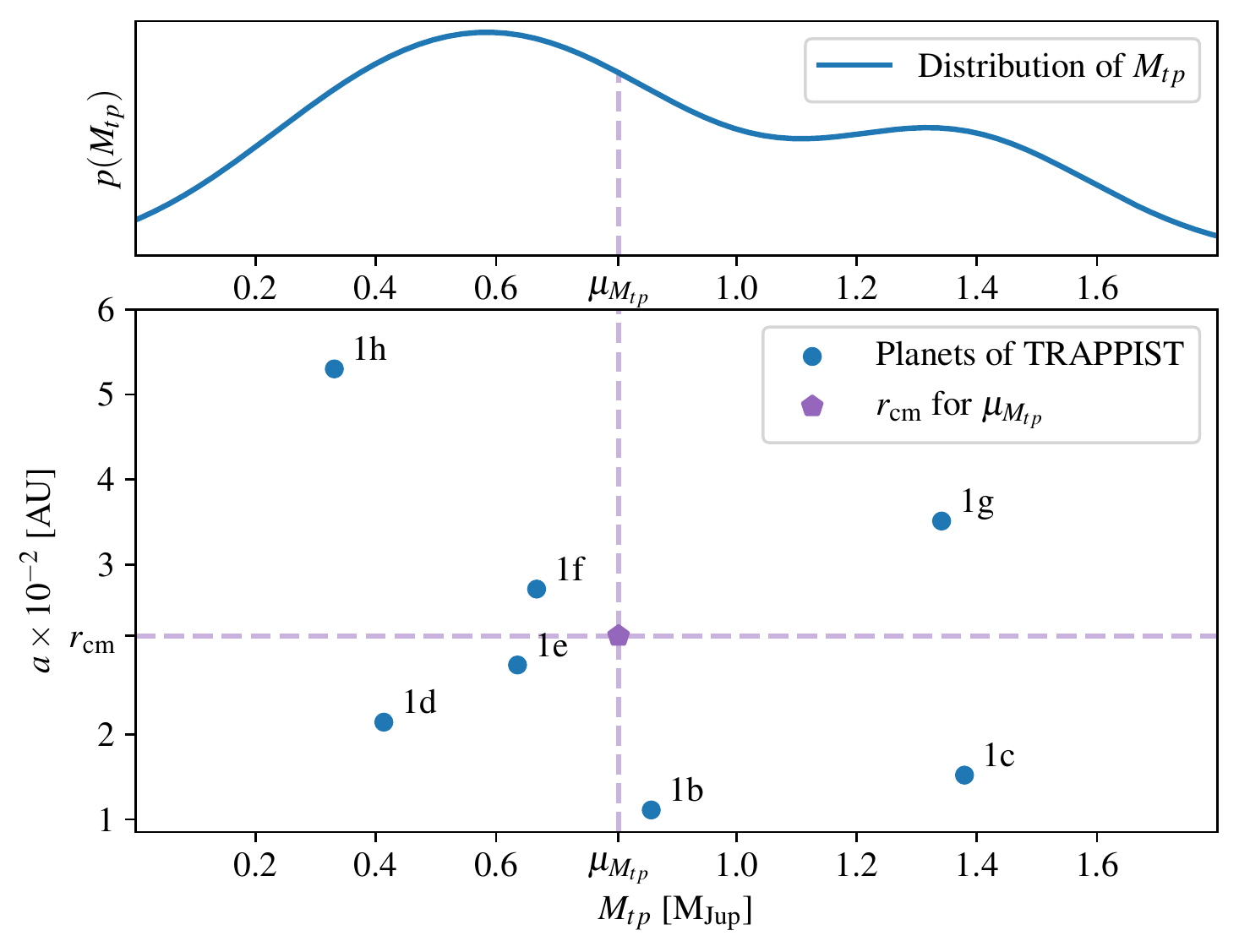}
    \caption{Kernel density estimation of the planet mass distribution $\mu_{M_\text{tp}}$ (above) and semi-major axis vs. planet mass (below) for the TRAPPIST-1 system. The dashed line corresponds to the first moment of the planet mass wrt the semi-major axis, or \emph{center of mass} $r_\text{cm}$.}
    \label{fig:CenterOfMass}
\end{figure}

\subsection{Synthetic data}\label{synth_data}
Planet population synthesis models describe how the physical evolution of protoplanetary disks results in planetary systems with properties that agree with observations. \cite{mordasini09_I, mordasini09_II, alibert11} and \cite{miguel11} inasmuch observed planetary systems that can be reproduced using these formulations. Current models strive to reach self-consistent, generative recipes for planetary formation with the use of Monte Carlo techniques (\citealp{Benz}), but observational biases create a gap between theoretical predictions and observations.

In order to create a synthetic planetary data set we follow the Monte Carlo planet formation recipe of \cite{miguel11,chaparro_2018}. Here, planets grow in individual systems following an oligarchic growth model of solid cores coupled with a core instability model for gas accretion along with Types I and II migration. The physical variables that we use as initial conditions for the simulations are summarized in Table \ref{Tab:initial_conditions}, along with the probability distributions from which the initial conditions for individual simulations were drawn. These distributions, which effectively work as prior probability distributions are constructed from observations of exoplanetary systems \citep{miguel11}.

The results of these simulations are masses and semi-major axes for all planets formed in individual systems. From this we compute physical properties that describe each system which allows us to make a statistical comparison to observed exoplanetary systems. 

We thus ran simulations yielding 1200 synthetic systems for which properties listed in Table~\ref{Tab:initial_conditions} are known. It is noteworthy that according to these planet population synthesis simulations only about 20\% of systems have giant planets.

\begin{center}
    \begin{table}
    \centering
    \caption{Initial conditions for our planet population synthesis model. The probability distribution functions are constructed from observations of exoplanetary systems \citep{miguel11}.}%\vspace*{.1cm}
    \begin{tabular}{lll} \hline
    %\multicolumn{1}{c} Property & Symbol & Distribution  &   Region \\ \hline \hline 
        Stellar mass  $M_{\star}$ [$M_\odot$]   	& $\log(\mathcal{U}(a_{M_\star},b_{M_\star}))$& \begin{tabular}[l]{@{}l@{}} $a_{M_\star}\sim0.7$,\\ $b_{M_\star}\sim 1.4$.\end{tabular} 	\\ %\midrule
        Mass of the disk $M_{d}$ [$M_\odot$]		& $\log(\mathcal{N}(\mu_{M_d},\sigma_{M_d}))$ & \begin{tabular}[l]{@{}l@{}} $\mu_{M_d}=-2.05 $,\\ $\sigma_{M_d}=0.85$.	\end{tabular}	\\ %\midrule
        Radial extent $a$  [AU]  		& $\log(\mathcal{N}(\mu_{a_c},\sigma_{a_c}))$ & \begin{tabular}[l]{@{}l@{}} $\mu_{a}=3.8$, \\ $\sigma_{a}=0.81$.  		\end{tabular} 	\\ %\midrule
        Stellar metallicity [Fe/H]		& $\mathcal{N}(\mu,\sigma)$              	  & \begin{tabular}[l]{@{}l@{}} $\mu=-0.02$, \\ $\sigma=0.22$.            	\end{tabular}   \\ %\midrule
        Time of gas dissipation $\tau_{g}$ [y]		& $\log(\mathcal{U}(a_{\tau_g},b_{\tau_g}))$  & \begin{tabular}[l]{@{}l@{}} $a_{\tau_g}\sim 10^{6}$, \\ $b_{\tau_g}\sim 10^{7}.$\end{tabular}  \\ \hline  %\hline 
    \end{tabular}
    \label{Tab:initial_conditions}
    \end{table}
    \end{center}

%-------------------------------------- Classification
\subsection{Center-of-mass clustering of planetary systems}\label{Class}

Seeking to improve upon existing exoplanetary system classification schemes, and wishing to assess the role of observational biases when comparing observations with planet population synthesis results, we built an unsupervised learning model based on the exoplanet mass distribution along the orbital plane of each system. Existing classification schemes are based on ad-hoc comparisons of exoplanetary systems and mostly look at superficial similarities between planetary systems \cite{miguel11, Benz}, and are understandably concerned with matching properties of observed and simulated systems in order to avoid the effects of observational biases. Our approach to classification of observed and synthetic planetary systems is based on Gaussian Mixture Models (GMM) which is an unsupervised learning  technique that requires no previous labeling and allow for the discovery of patterns in data (\citealp{VanderPlas2016}). For this we used the the \texttt{scikit-learn} package \cite{scikit}.

Our variable of interest for this classification scheme needed to be system-wide variable with information about where in the disk most of the planet mass is located. We therefore used the first moment of planet mass distribution with respect to the semi-major axis. We call this variable the \emph{center of mass} radius $r_\text{cm}$. As an illustration of how this works, Figure \ref{fig:CenterOfMass} shows the total planetary mass distribution $M_\text{tp}$, for the TRAPPIST-1 system and its center of mass $r_\text{cm}$, according to the public data from \cite{exoplanet}. We also developed our own model validation technique based on the principles of Approximate Bayesian Computation (ABC).

We will refer to this classification scheme again in Section~\ref{results} to study the conditions in which planetary systems may shift clusters when considering gravitational interactions between planets.

\subsubsection{Observational data classification}\label{clobs}
\begin{figure}
    \centering
	\includegraphics[width=.85\columnwidth]{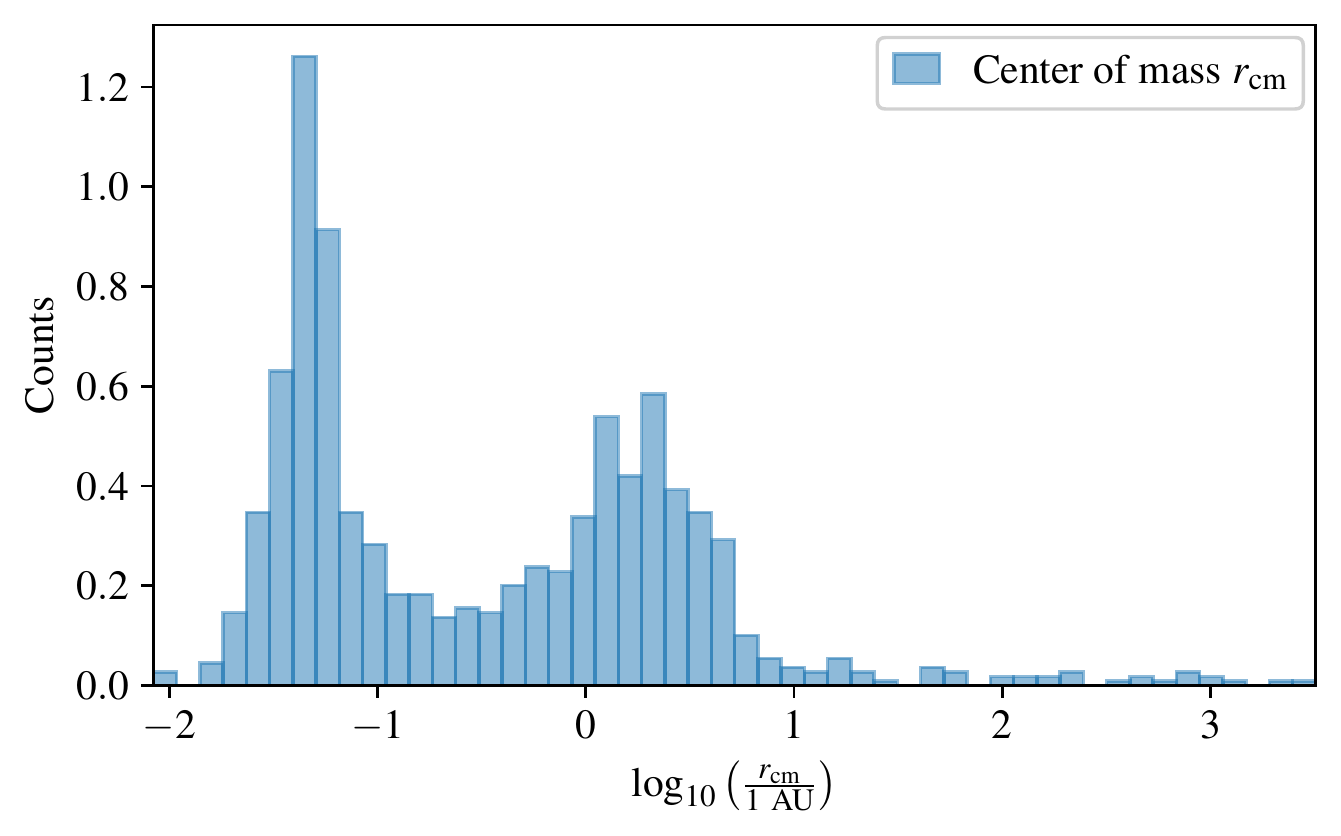}
    \caption{Histogram for the log-center of mass $r_\text{cm}$ of observed planetary systems. We use the log due to the fact that the center of mass range spans many orders of magnitudes.}
    \label{fig:hist_CoM}
\end{figure}

Figure \ref{fig:hist_CoM} shows the distribution of the log-center of mass $\log r_\text{cm}$ for observed exoplanetary systems. There is a region with a bi-modal behaviour between $10^{-2}\ \text{AU}<r_\text{cm}<10\ \text{AU}$: the first peak corresponding to systems with hot Jupiters, whereas the second peak corresponds to systems more alike to our own Solar System. Additionally, there is an uniformly distributed region at $r_\text{cm}>10$, corresponding to systems with planets which are located far away from their host star. In order to move beyond this superficial classification, we built an unsupervised learning classification scheme based on GMM.

Since GMMs allow as many Gaussian clusters as there are data points, overfitting is a possibility. However it can be avoided with the use of \textit{information criteria} based on Occam's razor, which penalize models that require too many clusters or in general too many fitting parameters (\citealp{claeskens_hjort_2008}). The most popular and versatile estimators are the \textit{Akaike information criterion - AIC} proposed by \cite{Akaike1974}, and the \textit{Bayesian information criterion - BIC} defined by \cite{schwarz1978}.

\begin{figure}
    \centering
	\includegraphics[width=.9\columnwidth]{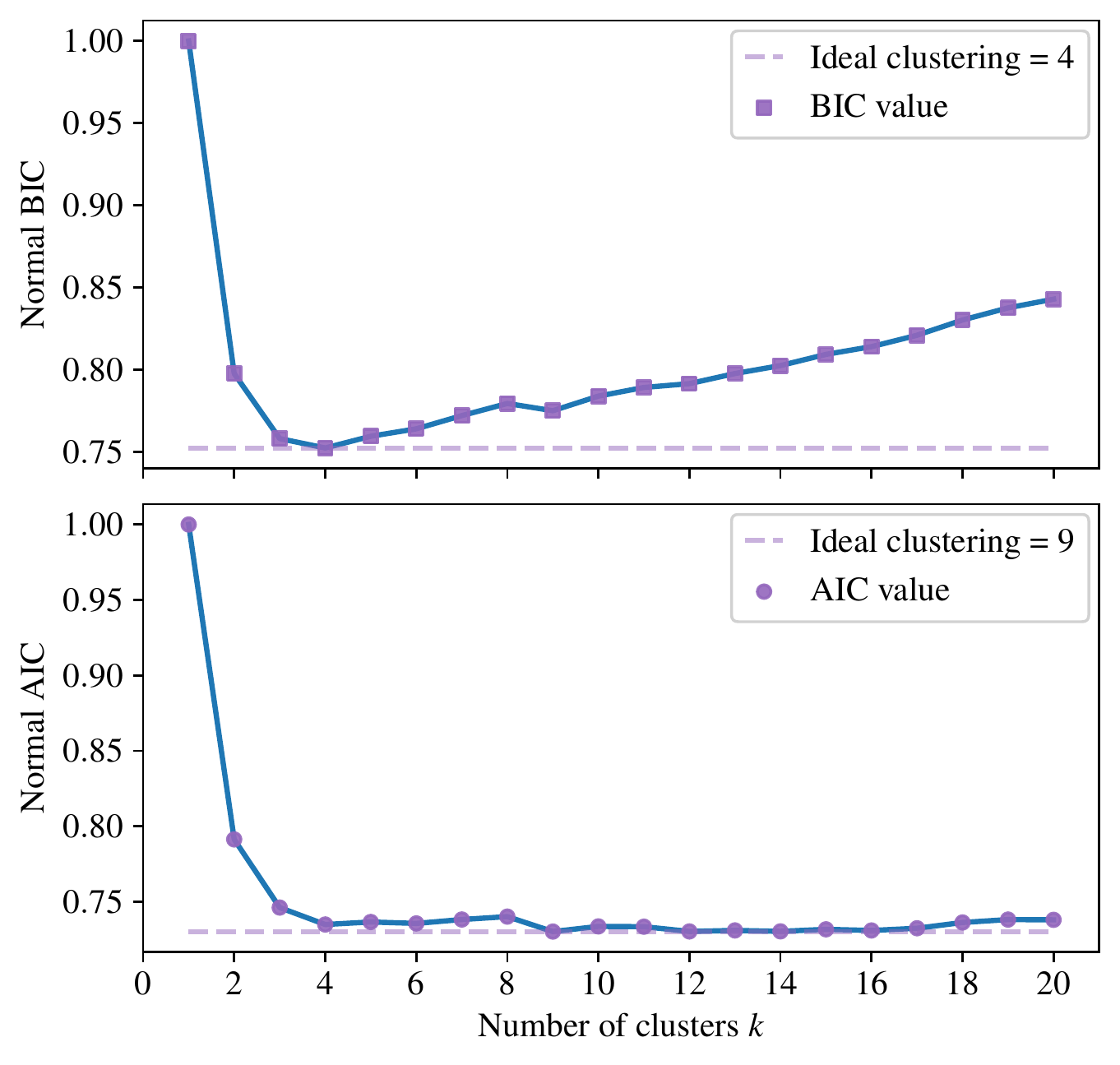}
     \caption{Normalized BIC and AIC values vs. number of GMM clusters $k$ for center-of-mass clustering of observed exoplanetary systems.}
     \label{fig:criteria}
\end{figure}

\begin{figure} 
    \centering
	\includegraphics[width=0.85\columnwidth]{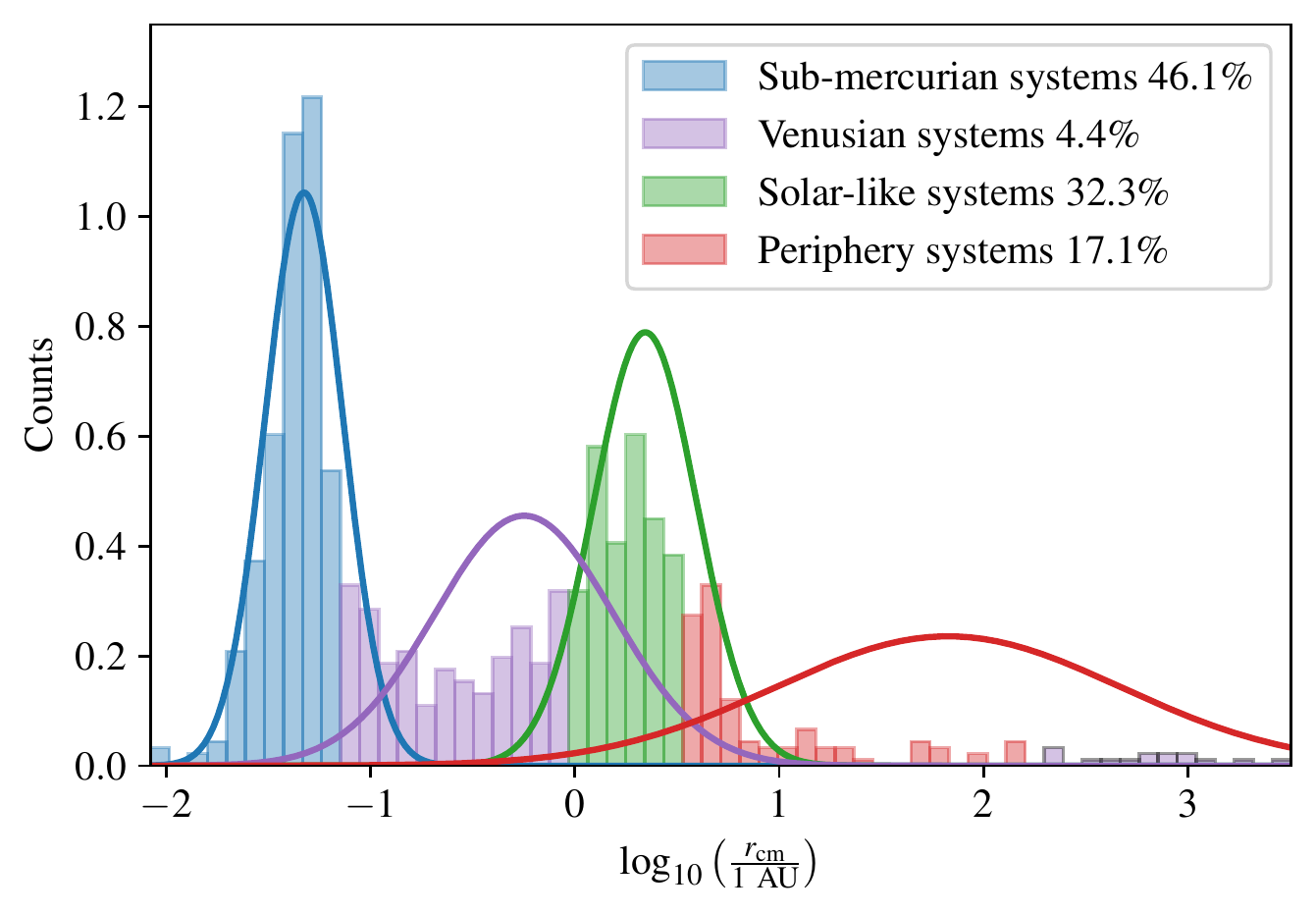}
    \caption{Gaussian Mixture Model with $k = 4$ clusters to classify observed exoplanetary systems according to their log center of mass.}
    \label{fig:class}
\end{figure}

We calculate these criteria for models with $k=1,2,3,4,\ldots,20$ clusters. The minimum result in the values of AIC and BIC indicates the optimal mixture model. Figure \ref{fig:criteria} shows normalized values of AIC and BIC for log-center of mass clustering. The lowest BIC value is for a model with $k=4$ clusters, and the lowest AIC value is for a $k=9$ clusters. Thus, the  BIC recommends a simpler model, which is preferred because BIC penalizes model complexity more heavily, as it accounts for the number of fitting parameters, while the AIC does not. Figure \ref{fig:class} shows the Gaussian Mixture Model with $k=4$ clusters applied to the observed center of mass $r_\text{cm}$.

For easy reference, we named clusters by matching their mean center of mass with objects in the the Solar System orbiting at similar distances as follows: \textit{Sub-Mercurian} ($0.01\ \text{AU}\leq r_\text{cm} < 0.1$ AU) are systems that have a center of mass close to their host star, \textit{Venusian} ($0.1\ \text{AU} \leq r_\text{cm} < 1$ AU) are systems with a center of mass comparable to the approximate distance from Venus or the Earth to the Sun, \textit{Solar-like} ($1\ \text{AU} \leq r_\text{cm} < 10$ AU) are planetary systems with their center of mass located at a distance similar to that of Jupiter, which holds most of the planetary mass in our own Solar System, and \textit{Peripheral} ($r_\text{cm}>10$ AU), systems whose center of mass is located beyond the regions where we find planets in our own Solar System. 
\subsubsection{Approximate Bayesian Computation validation}
The classification scheme presented above is based upon a frequentist approach that does not consider the measurement uncertainties (\citealp{ExEp}). In our case, the main sources of uncertainty is the planet mass and semi-major axis. Therefore, inspired by Approximate Bayesian Computation (ABC) we created a Monte Carlo validation scheme based on evaluating the robustness of the classification when checked against measurement errors. The approach is summarized as follows.

For each system we draw values of $r_\text{cm}$ from the posterior distribution generated from the center of mass estimation, taking the prior distributions for the planet mass and semi-major axis as Gaussians with mean and variance from the reported measurement and corresponding error. We then check whether for 1000 draws from the $r_\text{cm}$ posterior of each system, the posterior distribution of the clustering model predicts a different cluster from the original. Since we do not update the posterior distribution of the clustering model this validation scheme belongs to the ABC family of algorithms. If the posterior prediction re-classifies a system (with respect to the original classification) for more than 5\% of the draws, we consider this an \emph{in-betweener} system. Our validation scheme yields that less than $2\%$ of observed exoplanetary systems are \emph{in-betweeners}, which shows that our GMM-based classification scheme for $r_\text{cm}$ across systems is indeed robust. 

\subsubsection{Synthetic data classification}
\begin{figure} 
    \centering
	\includegraphics[width=0.9\columnwidth]{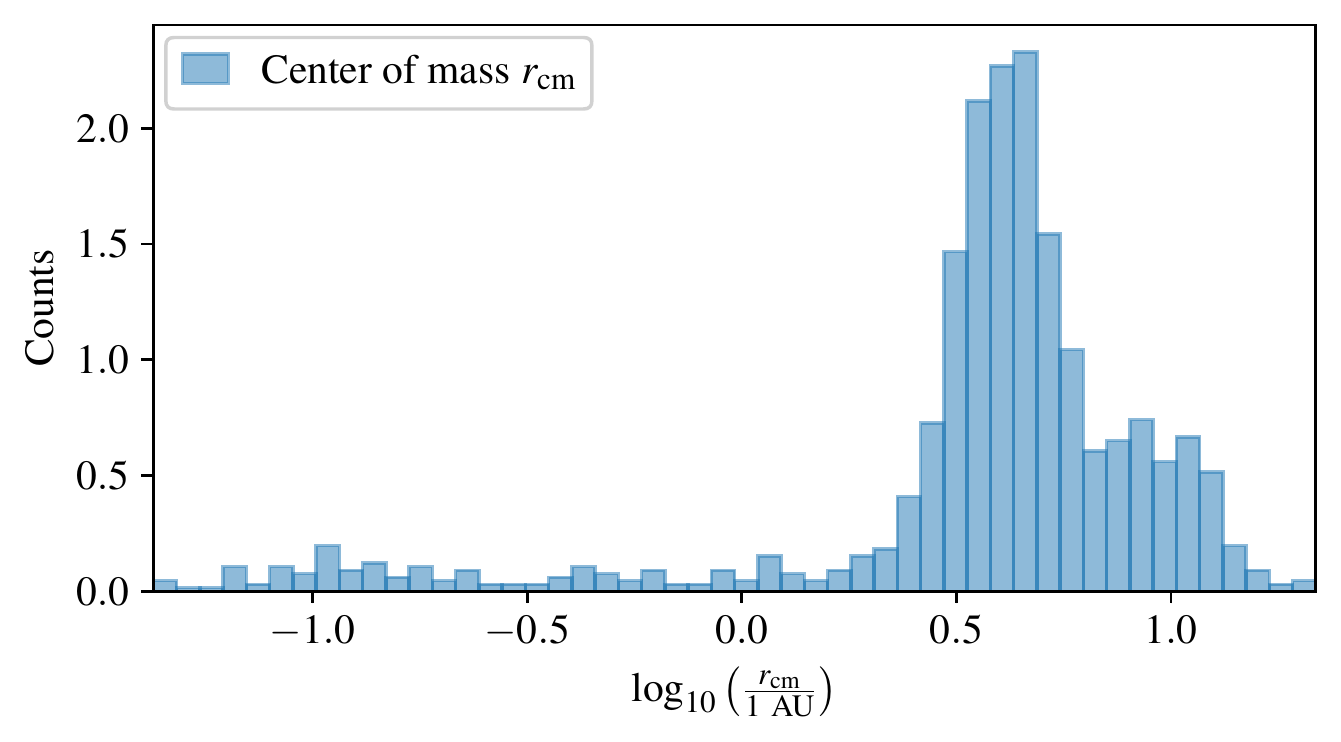}
    \caption{Log center of mass distribution for our synthetic planetary systems.}
    \label{fig:SCoM}
\end{figure}

\begin{figure} 
    \centering
	\includegraphics[width=0.9\columnwidth]{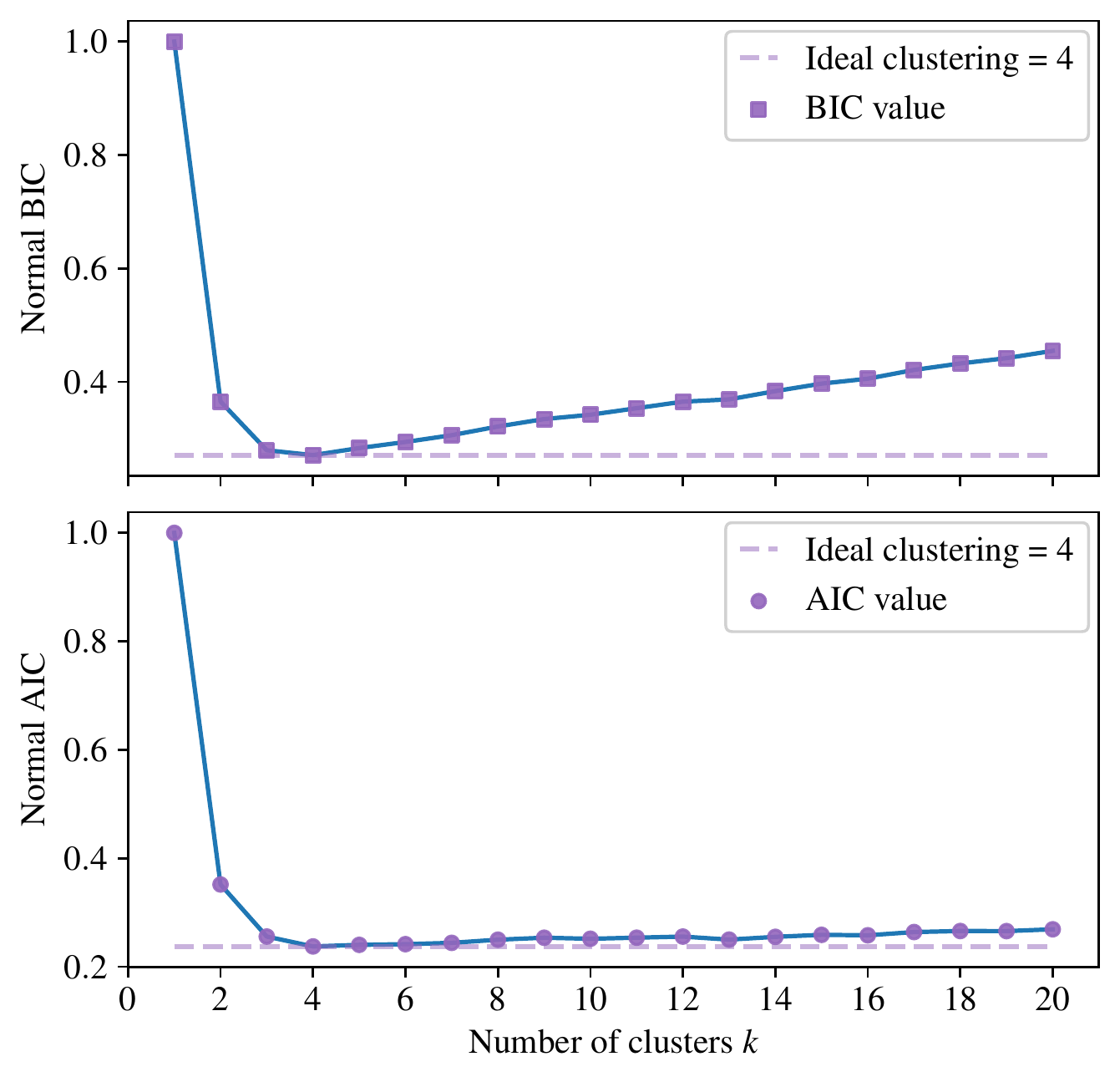}
    \caption{Normalized BIC and AIC values vs. number of GMM clusters $k$ for center-of-mass clustering of synthetic planetary systems.}
    \label{fig:criteria2}
\end{figure}

\begin{figure} 
    \centering
	\includegraphics[width=0.85\columnwidth]{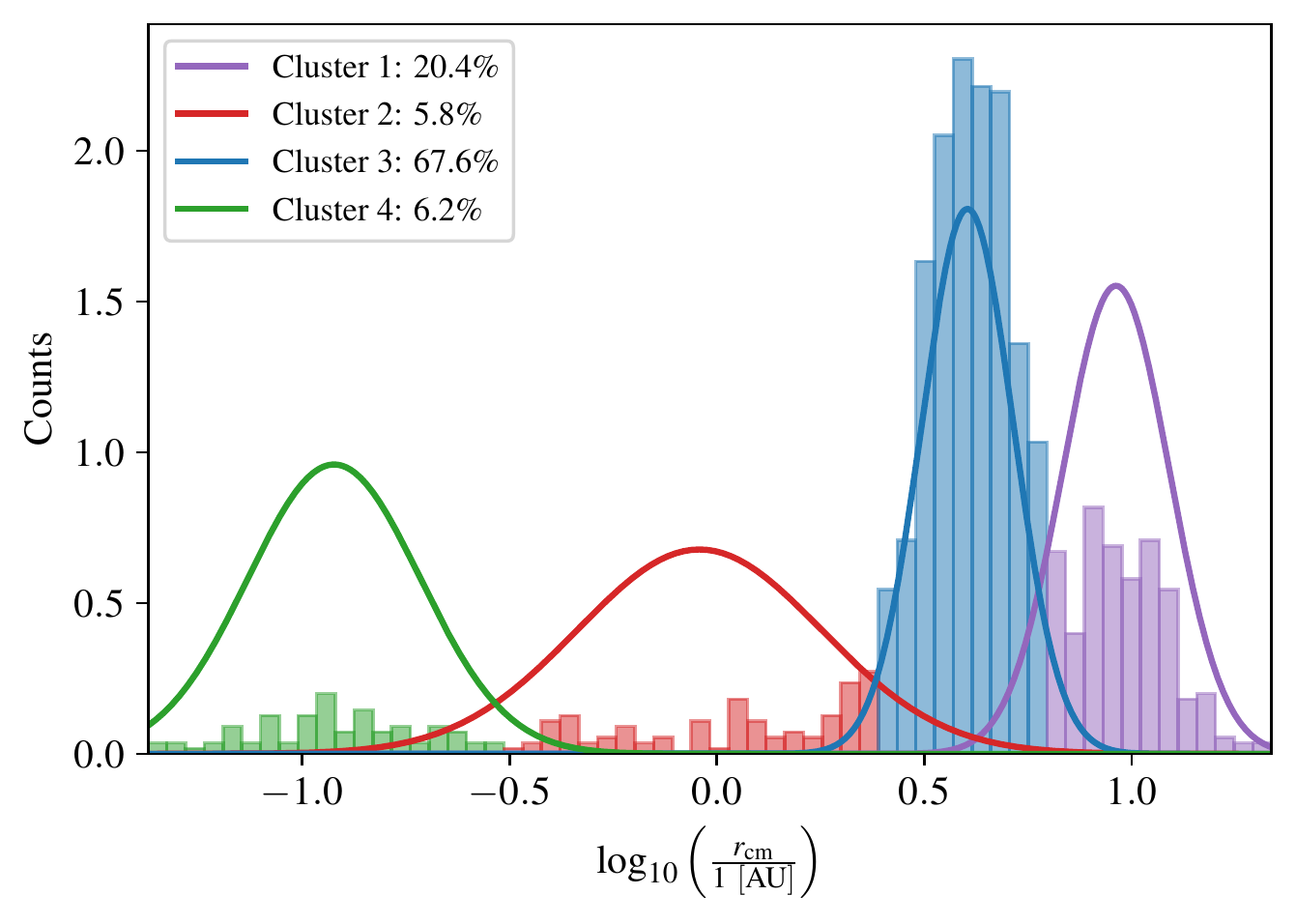}
    \caption{Gaussian Mixture Model with $k = 4$ clusters to classify synthetic planetary systems according to their log center of mass.}
    \label{fig:classif_sy}
\end{figure}

In order to better understand how observational biases may affect classification schemes, we applied the same classification scheme shown above to the log-$r_\text{cm}$ of synthetic systems generated from the planetary population synthesis model of \cite{miguel11,chaparro_2018}, described in Section~\ref{synth_data}. The center of mass histogram in Figure \ref{fig:SCoM} shows a uniform distribution of systems with center of mass between $0.1$ AU and $1.5$ AU, an approximately bi-modal distribution with a first peak near $5$ AU and a second peak near $10$ AU, and very few systems with centers of mass much greater than this value. This is because observed planetary systems with large centers of mass very likely have additional, yet unobserved planets close to their star, which would bring the value of $r_\text{cm}$ down after detection. A comparison between the frequency of Solar-like systems relative to Sub-Mercurian in observational (Figure~\ref{fig:class}) and synthetic (Figure~\ref{fig:SCoM}) data shows that current observations may be missing a factor of aboyt 20x Solar-like exoplanetary systems (e.g. with centers of mass near 1-6 AU), and an additional population of systems with centers of mass near 10 AU. A significant sample of these populations us still beyond our current detection capabilities. 
Figure \ref{fig:criteria2} shows the BIC and AIC for the GMM-based clasification of planetary systems which yields $k=4$ as the most reasonable number of clusters, shown in Figure \ref{fig:classif_sy}. The classification schemes for observational and synthetic data roughly agree in that there are four populations of systems (i.e. Sub-Mercurian, Venusian, Solar-like, Peripheral according to our classification in Section~\ref{clobs}), although our simulations show that the three clusters closest to the star are higher in center of mass by about half an order of magnitude with respect to the observational clusters. Again, this is explained by observationalal biases that favor detection of systems with massive, close-in planets, where further detections of high-mass, located at distances $\gtrsim1$ AU from their star are unlikely with our current technology. \\

On the other hand, observational data in Figure \ref{fig:class} shows a fraction of systems with very large centers of mass ($r_\text{cm}\gg10$ AU) that does not appear to be significant in synthetic systems (Figure \ref{fig:classif_sy}). In the clustering of synthetic systems, the \emph{peripheral} cluster is located between 6-20 AU, which is a significantly lower $r_\text{cm}$ range than what observations seem to show. Systems in this range have been found by techniques such as gravitational lensing and direct imaging, which are biased toward extremely far-away planets \citep{kipping16} with respect to other detection methods such as transit and radial velocity. Thus, these observed Periphery sytems are likely extreme $r_\text{cm}$ outliers which do not constitute a fair sample of all Periphery systems, most of which ($r_\text{cm}\sim10$ AU) would currently be beyond our detection capabilities.

%---------------------------------------------------------------------------------------------------------------------
\section{Tools for assessing the role of gravitational interactions}\label{methods}

In this section we discuss the gravitational-collision model and the statistical framework that will allows us to assess to what extent the final configuration of a synthetic planetary system is dependent on its secular evolution. To this end, we use an $N$-body numerical Leapfrog integrator of the equations of motion including gravitational interactions between all bodies. We also include a collision recipe based on the work of \citet{ALEXANDER1998113} which considers dynamic gravitational interactions between bodies with a variable time step for when collisions are in play. This formulation also includes recipes for rebound, fragmentation and mass exchange. This allows for an accurate tracking of planetary evolution.

\subsection{Collision model}\label{col_model}
The outcome of a collision depends on the impact energy vis-a-vis and the resistance or hardness of the bodies. We first consider the collision of two planetesimals with total mass $M = m_1 + m_2$ ($m_{1 2}$) and with $\textbf{v}_1$ and $\textbf{v}_2$ as their respective velocities.

The possible outcomes of collisions can be: i. rebound, ii. rebound with mass exchange and iii. fragmentation; these lead to accretion when the planetesimals are dynamically unexcited, i.e., low impact velocity. On the other hand, high velocity impacts in dynamically excited populations (the present-day asteroid belt, debris disks, etc.) lead to a sequence of collisions that grind the bodies into very small particles \cite{ARMITAGE2018}.

We define the total impact energy due to the relative motion of the bodies with respect to the center of mass as,

\begin{equation}\label{energy}
    E = \frac{1}{2} \frac{m_1 m_2}{m_1 + m_2} V^2_r\ .
\end{equation}

Here $\textbf{V}_r = \textbf{v}_2 - \textbf{v}_1$. Since these collisions are expected to be mostly inelastic, a fraction $\eta$ of the energy is dissipated as heat during the encounter. Thus, the energy $E'$ after a collision would be,

\begin{equation}
    E' = (1 - \eta)E\ .
\end{equation}

Equation \ref{energy} yields the \textit{rebound velocity}, i.e. the relative velocity between the two bodies immediately after the collision,

\begin{equation}
    \textbf{V}_\text{reb} = - c_i\textbf{V}_r\ .
\end{equation}
The quantity $c_i=\sqrt{1- \eta}$ is the restitution coefficient, which can be considered as the ratio of the rebound and impact speeds.

\subsubsection{Rebound}\label{rebound}
If the relative velocity is too low the collision will result in a rebound, i.e. both bodies collide and separate without exchange of mass. For this to occur the maximum stress on the surface of the body must be less than the impact strength of the planetesimal \citep{GREENBERG19781}. In other words, the relative speed must satisfy the following condition,

\begin{equation}
	V_r < V_c = \frac{2S}{c\rho_c} \ .
\end{equation}

Here $S$ is the impact strength defined as the capacity of a material to absorb energy and deform plastically without fracturing, $c$ is the speed of sound and $\rho_c$ is the characteristic density\footnote{We are currently working on a self-consistent treatment of body densities, but for the time being such a model is beyond the scope of this work.}, which is of the order of the density of a typical rocky body in the Solar System. In this work we assume that for planets whose density $\rho$ is lower than our assumed value (e.g. for gas giants) there is a characteristic impact radius at which the planet behaves as a solid body with density $\rho_c$.

The new velocities after the collision are then,

\begin{equation}
\begin{split}
	\mathrm{\mathbf{v}}'_1 &= \mathrm{\mathbf{V}}_G - \left(\frac{m_2}{M}\right)\mathrm{\mathbf{V}}_\text{reb}\ , \\ \\
	\mathrm{\mathbf{v}}'_2 &= \mathrm{\mathbf{V}}_G + \left(\frac{m_1}{M}\right)\mathrm{\mathbf{V}}_\text{reb}\ .
\end{split}
\end{equation}

Here $\mathbf{V}_G$ is the velocity of the center of mass of the two colliding bodies.

\subsubsection{Rebound with mass exchange}\label{reb_crat}

On the other hand, if $V_r > V_c$ but the impact energy is not high enough to destroy the bodies, mass exchange occurs and the bodies will drift apart with new velocities that are a fraction of the initial impact velocity. The mass exchange process has two consequences: the loss of a small amount of mass and modification of the coefficient of restitution, i.e., the rebound velocity-impact ratio is reduced because there is a greater loss of energy in the impact. The modified coefficient of restitution is $c_{ii}$.

To account for this event, we need to calculate the mass loss and the velocity of the ejected debris. The ejected mass from $m_1$ after a mass-exchange impact by a body with mass $m_2$ with impact velocity $V_r$ can be approximated by \citep{GREENBERG19781},

\begin{equation}
	m_\text{ej}^{(1)} = c_\text{ej} K m_2 V_r^2 [V_e^{(1)}]^{-9/4}\ ,
\end{equation}

Here $K$ is the mass excavation coefficient, $c_\text{ej}$ is the ejecta velocity coefficient (both of which depend on intrinsic properties of the colliding bodies, see Table~\ref{tab:param_col}), and $V_e^{(1)}$ is the escape speed of the body with mass $m_1$. In analogy, this analysis works for $m_2$ as well. 

Considering mass exchange, the new masses are then,
\begin{equation}
\begin{split}
	m_1' &= m_1 - m_\text{ej}^{(1)} + m_\text{ej}^{(2)}\ , \\
	m_2' &= m_2 - m_\text{ej}^{(2)} + m_\text{ej}^{(1)}\ . \\
\end{split}
\end{equation}
The new velocities can be derived to be,
\begin{equation}
\begin{split}	
	\mathrm{\mathbf{v}}'_1 &= \mathrm{\mathbf{V}}_G - \left(\frac{m_2}{M}\right)\mathrm{\mathbf{V}}'_\text{reb}\ , \\ \\
	\mathrm{\mathbf{v}}'_2 &= \mathrm{\mathbf{V}}_G + \left(\frac{m_1}{M}\right)\mathrm{\mathbf{V}}'_\text{reb} \ .
\end{split}	
\end{equation}

\subsubsection{Fragmentation}\label{fragmentation}
If the impact energy $E$ is large enough, the internal structure of the body is destroyed. Fragmentation occurs when the energy per unit volume is greater than the impact strength, i.e. when $E > SW$, with $W$ being the volume of the planet, the body will be shattered. After such a collision, fragments may coalesce again and new bodies may emerge.

The fragments of the destroyed body follow a power-law mass distribution \citep{GREENBERG19781}, which yields fragments of which the largest has a mass $m_\text{max}$ that from experiments \citep{FUJIWARA1977277} can be calculated to be,
\begin{equation}
	m_\text{max} = \frac{M}{2} \left( \frac{E}{SW} \right)^{-1.24}\ .
\end{equation}
Here $M = m_1+ m_2$.

Following Section~\ref{reb_crat}, the amount of mass leaving the two-body system is,
\begin{equation}
	m_\text{ej} = c_\text{ej} M (V_e)^{-9/4}\ .
\end{equation}
Here $V_e$ is the escape speed for $M$. The remaining mass will coalesce into a new single body. For the mass distribution of the fragments escaping from the parent body a small number of bodies represented by four escaping bodies is assumed; it was chosen this way in order to prevent the number of particles from becoming too large. The mass of the new fragments is then,

\begin{align}
	\begin{drcases}
		M_1 = \frac{m_\text{ej} m_\text{max}}{M}\ , \\
		M_2 = \frac{1}{\alpha}m_\text{max}\ , \\ 
		M_3 = M_4 = \frac{3 - \alpha}{2\alpha} m_\text{max}\ , \qquad
	\end{drcases}	
	\; \alpha \geq 1 \\
	\begin{drcases}
		M_1 = M_2 = M_3 = M_4 = \frac{m_\text{ej}}{4}\ , \qquad
	\end{drcases}
	\; \alpha < 1\ .
\end{align}

Here $\alpha = 4 m_\text{max}/M$ is a proxy for the impact energy ($\alpha = 4 m_\text{max}/M = 2(E/SW)^{-1.24}$). Table \ref{tab:param_col} summarizes the numerical values chosen for the collision parameters.

\begin{table}
	\centering
	\caption{Collision parameters (SI units) following \protect\cite{ALEXANDER1998113}.}
	\label{tab:param_col}
	\begin{tabular}{lccc} % four columns, alignment for each
	Parameter & Symbol & Value & Units\\
		\hline
		Restitution coefficient          & $c_i$    & $0.7$               & -\\
		Modified restitution coefficient & $c_{ii}$ & $0.5$               & -\\
        Maximum rebound velocity         & $V_c$    & $55$                & $\si{m/s}$              \\
        Mass excavated coefficient       & $K$        & $1.0 \times10^{-7}$ & $\si{1/(kg.m^2/s^2)}$  \\
        Impact strength                  & $S$        & $3.0 \times10^7$    & $\si{J/m^3}$            \\
        Ejecta velocity coefficient      & $c_\text{ej}$ & $3.0 \times10^6$    & $\si{m/s}$          \\
        Characteristic density                          & $\rho_c$   & $3 \times 10^3$     & $\si{kg/m^3}$    \\
		\hline
	\end{tabular}
\end{table}

\subsection{Statistical framework}\label{stats_frame}
In order to make a global assessment of how a planetary system changes after considering its secular evolution, we built a statistical framework that allows us to directly compare how the planet architecture changes in a system. To do this, we non-parametrically estimate a probability distribution function for the planet eccentricities, masses, and semi-major axes for each system, before and after the planet orbits were allowed to evolve dynamically.

We estimate these distributions with the Kernel Density Estimation (KDE) method. For a random sampling $X_i$ of a given variable $x$, the KDE probability distribution function is,

\begin{equation}
	p_\text{KDE}(x) = \frac{1}{nh}\sum_{i}K\left( \frac{x - X_i}{h} \right)\ .
\end{equation}

Here $K$ is a smooth function called \textit{kernel}, which we take to be Gaussian, and $h$ is a parameter known as the bandwidth, which controls the amount of smoothing. We robustly estimate this parameter using a cross-validation method instead of the usual rules-of-thumb that are found in the literature.

The divergence, or statistical "distance" between two distributions is a measure of how much one distribution function differs from another. In this work we will use the Kullback-Leibler divergence \cite{kullback1951information}, which is based on the Shannon information entropy, which in turn is inspired by the physical concept of entropy. 

The relative entropy or the Kullback-Leibler distance between two probability functions $p(x)$ and $q(x)$ is thus defined as,

\begin{equation}\label{eq_kl}
D(p||q) = \sum_{i} p(x_i)\log{\frac{p(x_i)}{q(x_i)}}\ .
\end{equation}

Here $||$ indicates "divergence". The KL divergence estimates the number of additional bits (i.e., computed with the base 2 logarithm) needed to represent an event of a random variable. The better the approximation, the less additional information is required. If the divergence KL is zero it means that the two distributions are identical. 

The Kullback-Leibler divergence (KL divergence hereafter) stands out among other statistical tests that measure the statistical distance between two distributions, such as the Kolmogorov-Smirnov test (KS) or the Cramer-von Mises criterion (CM), because it considers most of the domain of the distributions instead of focusing on point-estimates, thus giving equal weight to all the available data. The KL divergence is therefore sensitive to subtle differences near the tails of the distributions, which other frequentist tests tend to ignore.

\begin{figure}\label{ex_dist}
	\includegraphics[width=\columnwidth]{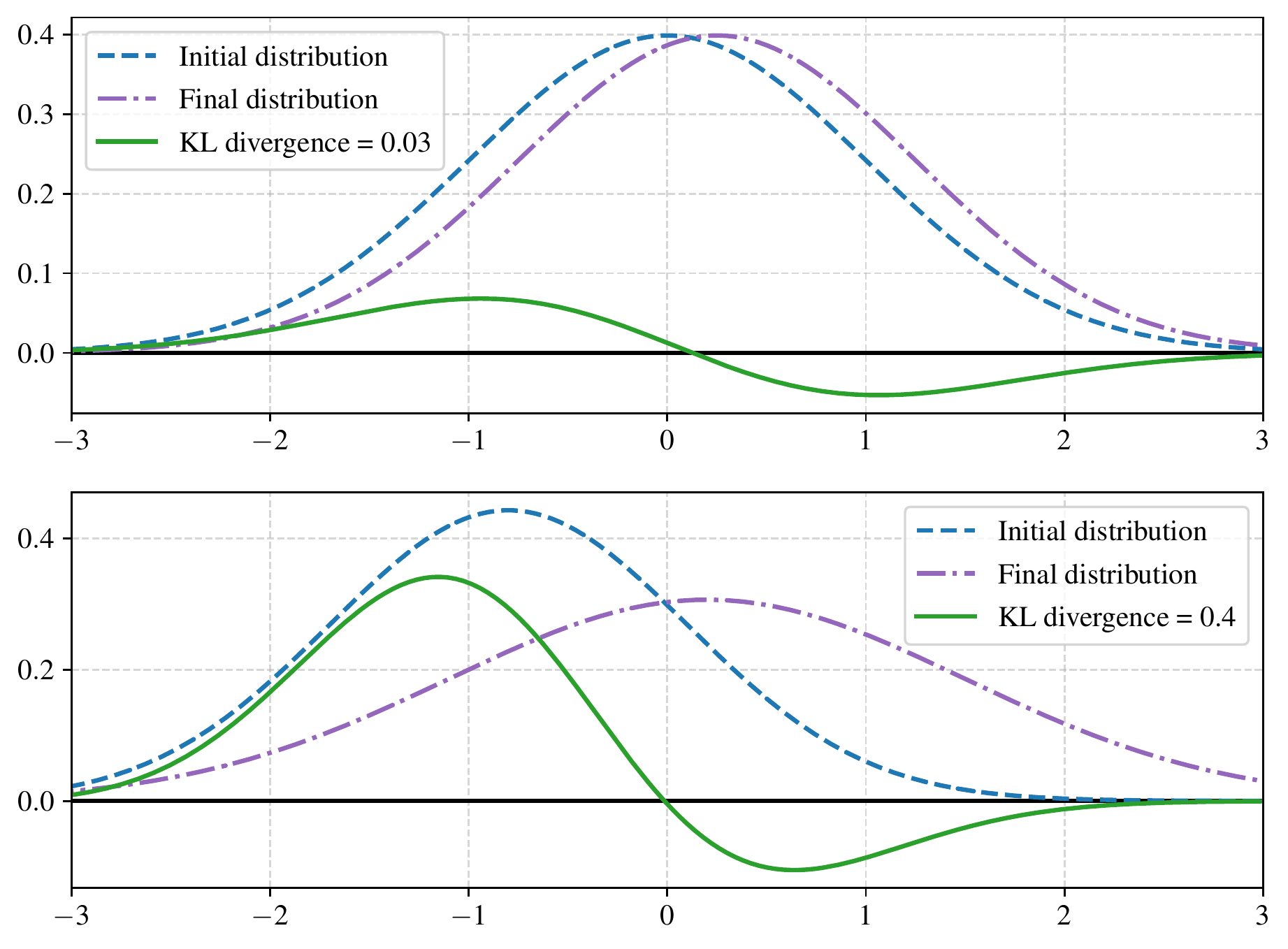}
    \caption{Illustrative plot showing how the Kullback-Leibler divergence works.}
    \label{fig:div_kl}
\end{figure}

Thus, we can use the KL divergence as an assessment of the change of the configuration of a planetary system before and after considering secular evolution for the selected variables: a significant change will yield a high KL divergence value and a small (or no) change will yield a KL divergence that is close to zero (see Figure~\ref{fig:div_kl} for an illustration). A systematic estimation of these changes across all synthetic systems from our Monte Carlo simulations will yield the parameter space of stellar parameters (e.g. stellar mass, metallicity) which will most likely result in a significant change in orbital configuration in terms of planetary mass, eccentricity, and semi-major axis.

%-------------------------------------- 

\section{Results}\label{results}
In this work we want to address the question of the role of secular gravitational evolution in the final configuration of planetary systems. Our approach is based on a statistical assessment based on comparing the initial and final distribution of planets using the statistical framework described in Section~\ref{stats_frame}. Additionally, we are interested in understanding how stellar parameters can have an influence in the dynamical evolution of the system.

We evolve synthetic planetary systems that are produced by a Monte Carlo planet population synthesis model (Section~\ref{synth_data} considering gravitational interactions (and their consequences) with an $N$-body collisional simulator. To this end we numerically solve the equations of motion for a given planetary system using a Leapfrog symplectic integrator. We chose this integrator because it ensures the energy conservation in large ($\sim10^6$~yr) time scales. Additionally, we use the collisional recipe described in Section~\ref{col_model} in order to account for rebound, fragmentation, and mass transfer between planets. 

We first ran a set of gravitational evolution simulations of a random sub-sample of 180 planetary systems over a period of $10^5$ years, and then a different group of 192 systems over $10^6$ years, adding up to a total $\sim4800$ planets. The physical parameters from which planetary systems were synthesized according to the planet formation recipe used here are summarized in \ref{Tab:initial_conditions}. The total computing time it took to run the simulations was $\sim64000$ h, running at the ClusterCIEN cluster of the Institute of Physics of Universidad de Antioquia.

The final step in preparing the data for a statistical comparison was to reconstruct the distribution of planetary mass, semi-major axis and eccentricity using a cross-validated Gaussian KDE for the systems in their initial and in their final configuration (i.e. after gravitational evolution).

\subsection{Secular evolution over $10^5$ yr}\label{ini_simulations}

\begin{figure}
	\centering
	\includegraphics[width=\columnwidth]{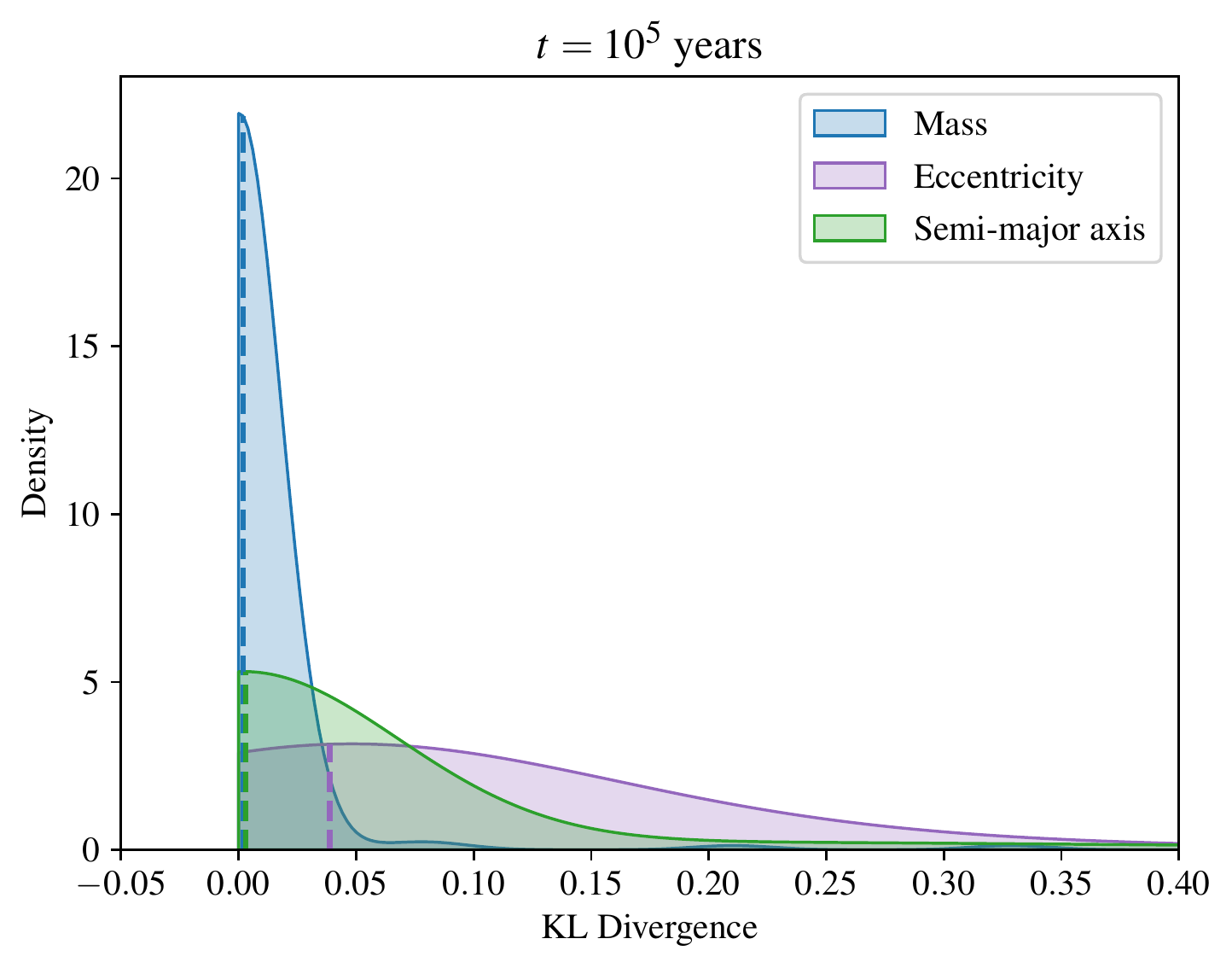}
	\caption{Gaussian KDE of the KL divergence between the mass, eccentricity and semi-major axis distributions of 180 planetary systems at $t=10^5$ yr.}
	\label{fig:div5}
\end{figure}

Figure~\ref{fig:div5} shows a Gaussian KDE of the KL divergence between the before-and-after gravitational evolution of 180 planetary systems in terms of their planet mass, eccentricity and semi-major axis distributions. Most of the systems show a maximum a posteriori (MAP, e.g. the mode) of the KL divergence distribution near zero, which means that most systems experience little-to-no change after considering the effects of gravitational interactions after $10^5$ yr. It also follows that the stability of a system on timescales up to $10^5$ yr does not depend on stellar parameters.

However, the MAP of the KL divergence distribution of eccentricity is higher than the MAP for planet mass and semi-major axis. This shows that systems tend to change their eccentricity more than their mass and semi-major axis configurations. Indeed, almost all of the systems in our sample show a more significant change in their eccentricities than in their mass and semi-major axis configurations. 

\begin{figure}
	\centering
	\includegraphics[width=\columnwidth]{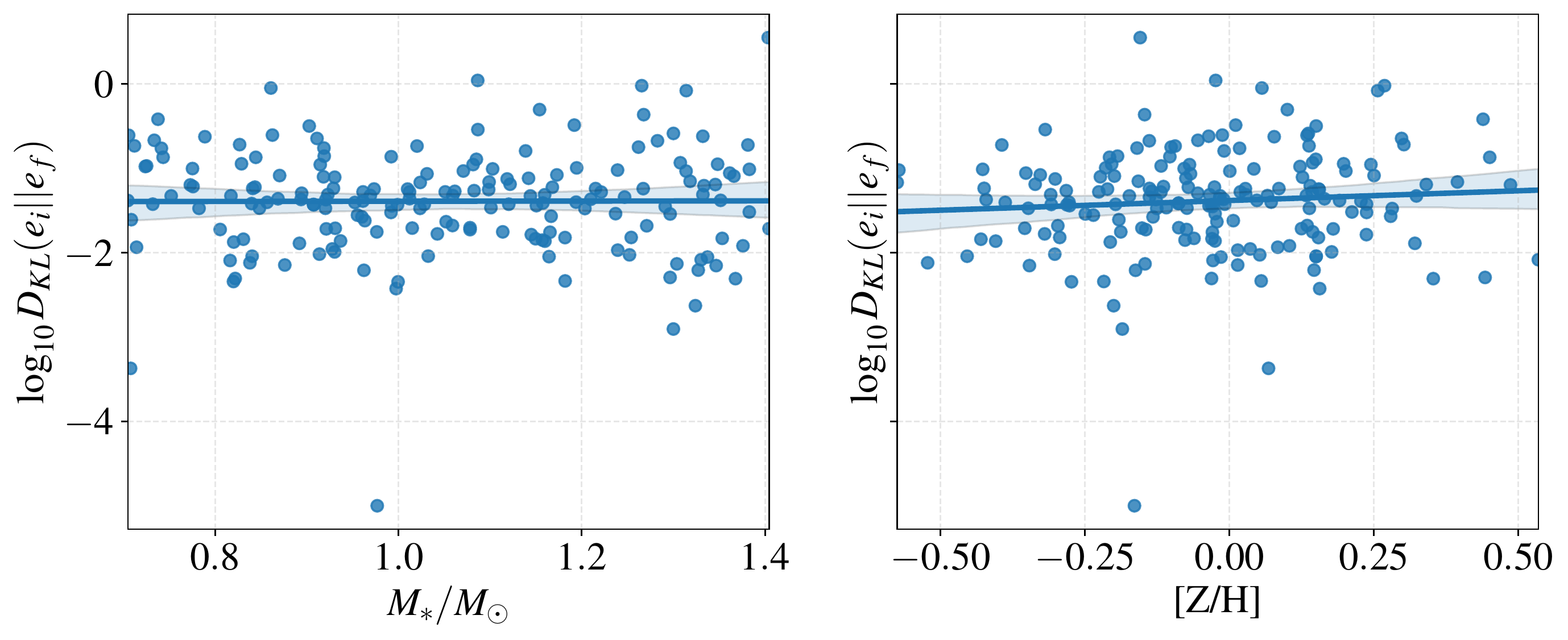}
	\caption{Linear regression of the KL divergence between the initial and final eccentricities vs. the mass and metallicity of the host star for 180 synthetic planetary systems at $t = 10^5$ yr. }
	\label{fig:ecc5}
\end{figure}

An inspection of the dependence of the eccentricity KL divergence on stellar parameters (Figure~\ref{fig:ecc5}) shows that most systems uniformly change their eccentricity configurations regardless of the stellar mass and metallicity, as there is virtually no correlation with those stellar parameters. These changes seem to arise normally due to gravitational interactions without radically changing the overall orbital configuration.

Figure~\ref{fig:div5} also shows the presence of extended tails in the KL distributions for planet mass and semi-major axis. We identify the source of this feature as \emph{outlier} systems, defined here as systems whose whose KL divergence is more than $1\sigma$ above the rest of the divergences from their sub-sample. Since one of our goals is to identify which factors may affect significant changes in the orbital configurations after evolution, we will discuss whether these outliers  emerge randomly or from a specific set of physical conditions in Section~\ref{marginal_cases}. The fraction of outliers in terms of planet mass and semi-major axis KL divergence is shown in Table~\ref{tab:outliers_systems_percentage}.

\subsection{Secular evolution over $10^6$ yr}\label{final_simulations}

\begin{figure}
	\centering
	\includegraphics[width=\columnwidth]{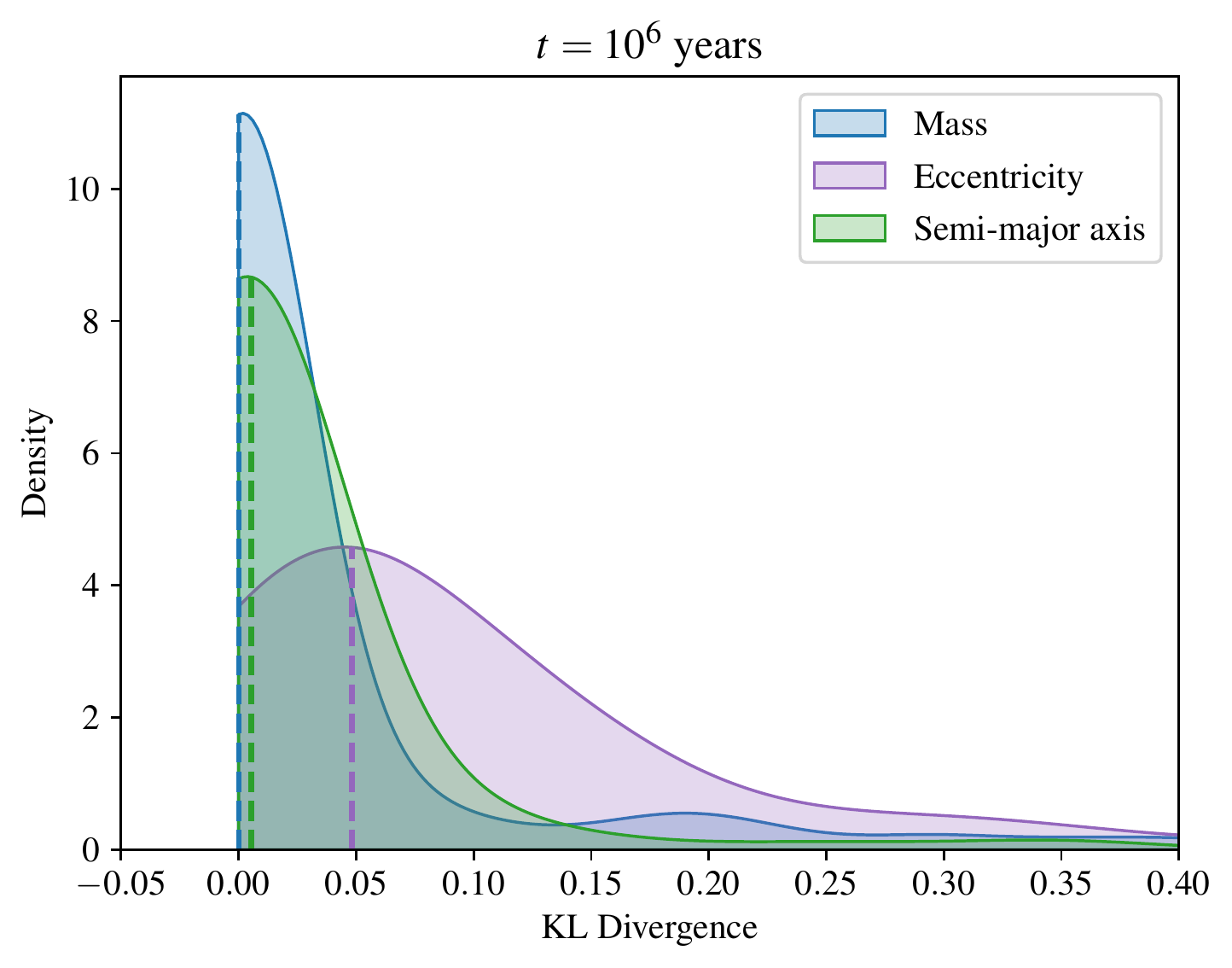}
	\caption{Gaussian KDE of the KL divergence between the mass, eccentricity and semi-major axis distributions of 180 planetary systems at $t=10^6$ yr.}
	\label{fig:div6}
\end{figure}

In order to achieve a realistic timescale for when a final configuration of a planetary system emerges, we took a sub-sample of 192 synthetic planetary systems and ran our gravitational collisional $N$-body simulations for 10$^6$ yr. Figure~\ref{fig:div6} shows a Gaussian KDE of the KL divergence between the initial and final configuration of planetary systems in terms of their orbital parameters. Similar to the $10^5$ yr-evolution case described in the previous section, the MAP of the KL divergence distributions for mass and semi-major axis are near zero, with a higher MAP of the KL divergence for eccentricity. Again, this means that a) most of the systems do not show a change in their planet mass and eccentricity configurations, whereas many systems change their eccentricities after $10^6$ yr of gravitational interactions, and b) that the stability of most systems for up to $10^6$ yr is not dependent on stellar parameters.  

Figures~\ref{fig:div5} and \ref{fig:div6} show that more systems evolving for 1 Myr change their orbital configuration when compared to systems evolving for only 100 kyr. Indeed, the fraction of systems that show a significant change in their planet mass and semi-mayor axis configurations in Table~\ref{tab:outliers_systems_percentage} shows this. Still, note that the fraction of systems that change their orbital configuration remains small.

\begin{figure}
	\centering
	\includegraphics[width=\columnwidth]{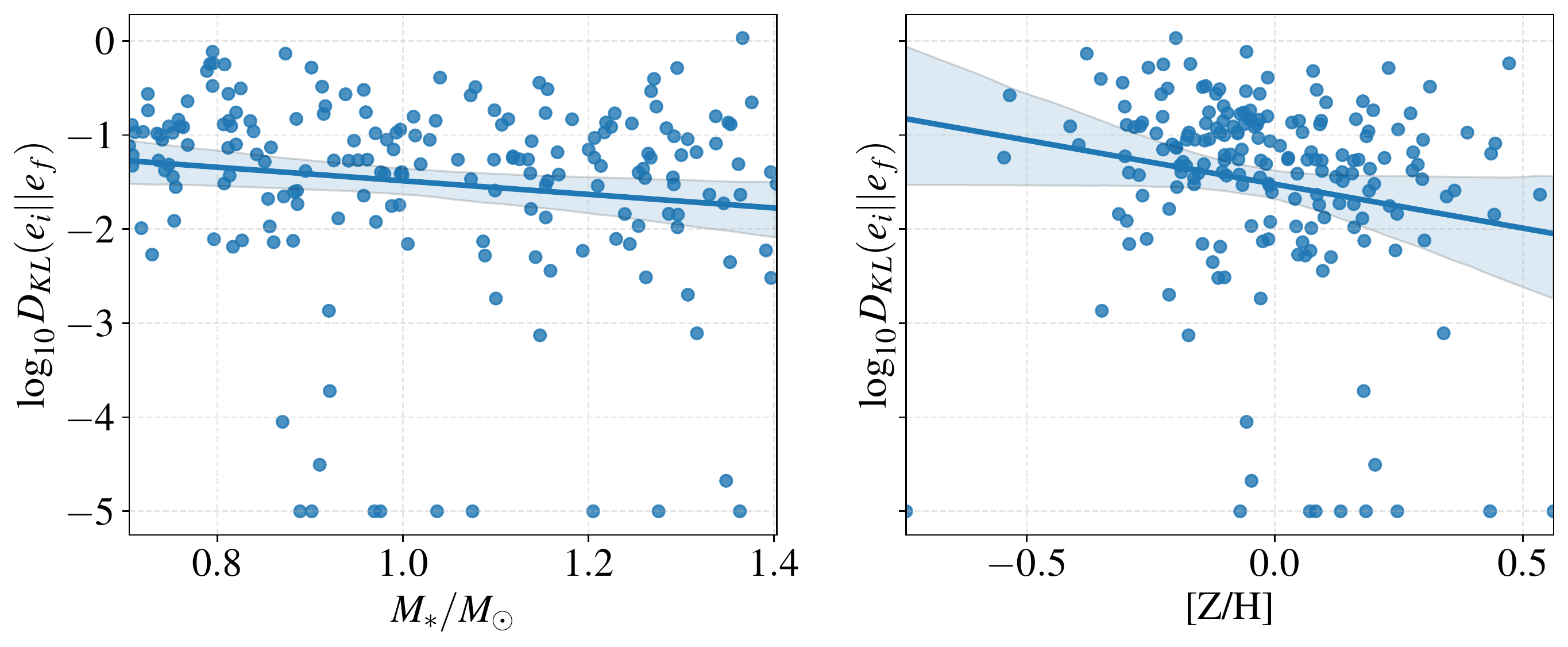}
	\caption{Linear regression of the KL divergence between the initial and final eccentricities vs. the mass and metallicity of the host star for 180 synthetic planetary systems at $t = 10^6$ yr. }
	\label{fig:ecc6}
\end{figure}

Figure~\ref{fig:ecc6} shows the eccentricity KL divergence vs. stellar parameters. Although there seems to be a negative correlationm the confidence bands in a linear regression show that this correlation is very weak (i.e. a slope = 0 model fits within the bands). Since there seem to be no particular outliers (other than at the low KL-divergence end) for either of the time periods considered in this work, we can conclude that even though planetary eccentricities are sensitive to gravitational interactions, they do not radically alter the overall planet mass or semi-major axis configurations.

\begin{table}
	\centering
	\caption{Percentage of outliers (i.e. systems changing their orbital configuration after gravitational evolution) based on the KL divergences between masses and semi-major axis distributions for $10^5$ and $10^6$ years of evolution.}
	\label{tab:outliers_systems_percentage}
	\begin{tabular}{ccc}
		\hline
		{} & \multicolumn{2}{c}{KL Divergence of:} \\
		\hline
		& Mass ($M_{\oplus}$)   & Semi-major axis (AU) \\
		\hline
		\% in sim: $10^5$ years & $2.77$       & $10.00$   \\
		\% in sim: $10^6$ years & $11.45$      & $5.72$    \\
		\hline
	\end{tabular}
\end{table}

\subsection{Outlier systems}\label{marginal_cases}

We now turn to a discussion on whether systems that experience changes in orbital configuration (outliers) as shown by an above $1\sigma$ deviation in their KL divergence appear at random or if they are more likely to appear in systems with specific ranges of stellar parameters. Table \ref{tab:outliers_systems_percentage} shows that less than 10\% of systems experience this change, i.e. about 40 systems.

We can therefore retrieve plausible correlations between stellar parameters and the likelihood that a planetary system will experience significant changes in its final orbital configuration. Given that most (90\%) of the systems in our sample do not experience such a change, i.e. their KL divergence remains near zero after gravitational evolution, we are interested in the few outlier systems that do experience changes in their orbital configuration, informed by the KL divergence. Recall that since we calculate different KL divergences for planet mass and semi-major axis distributions, there are \emph{mass} outliers and \emph{semi-major axis} outliers.

\begin{figure*}
    \centering
    \includegraphics[width=0.6\textwidth]{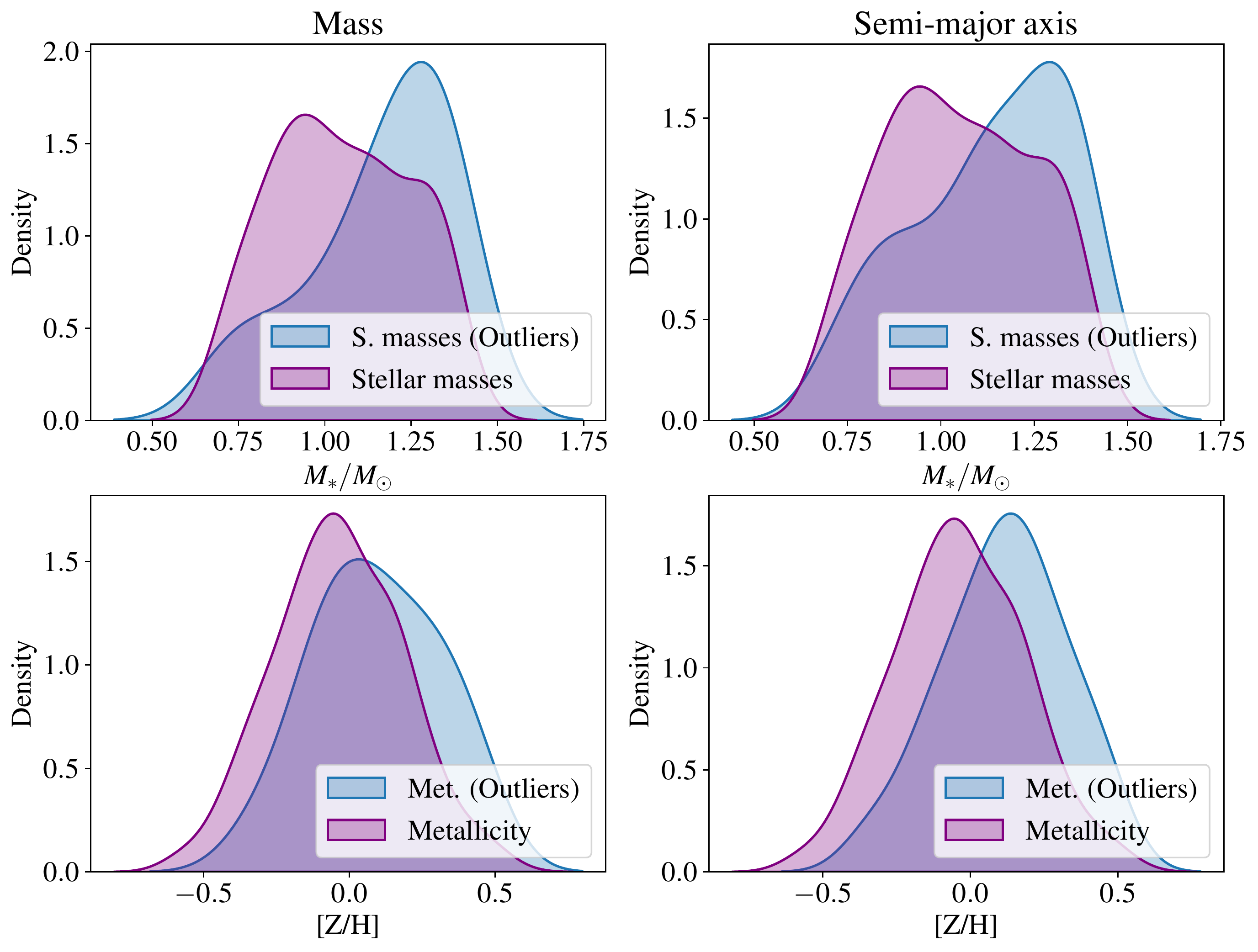}
    \caption{Stellar mass and metallicity distributions of the mass and semi-major axis outlier systems in the sim group: $10^5$ years.}
    \label{fig:st_out_1e5}
\end{figure*}

\begin{figure*}
    \centering
    \includegraphics[width=0.6\textwidth]{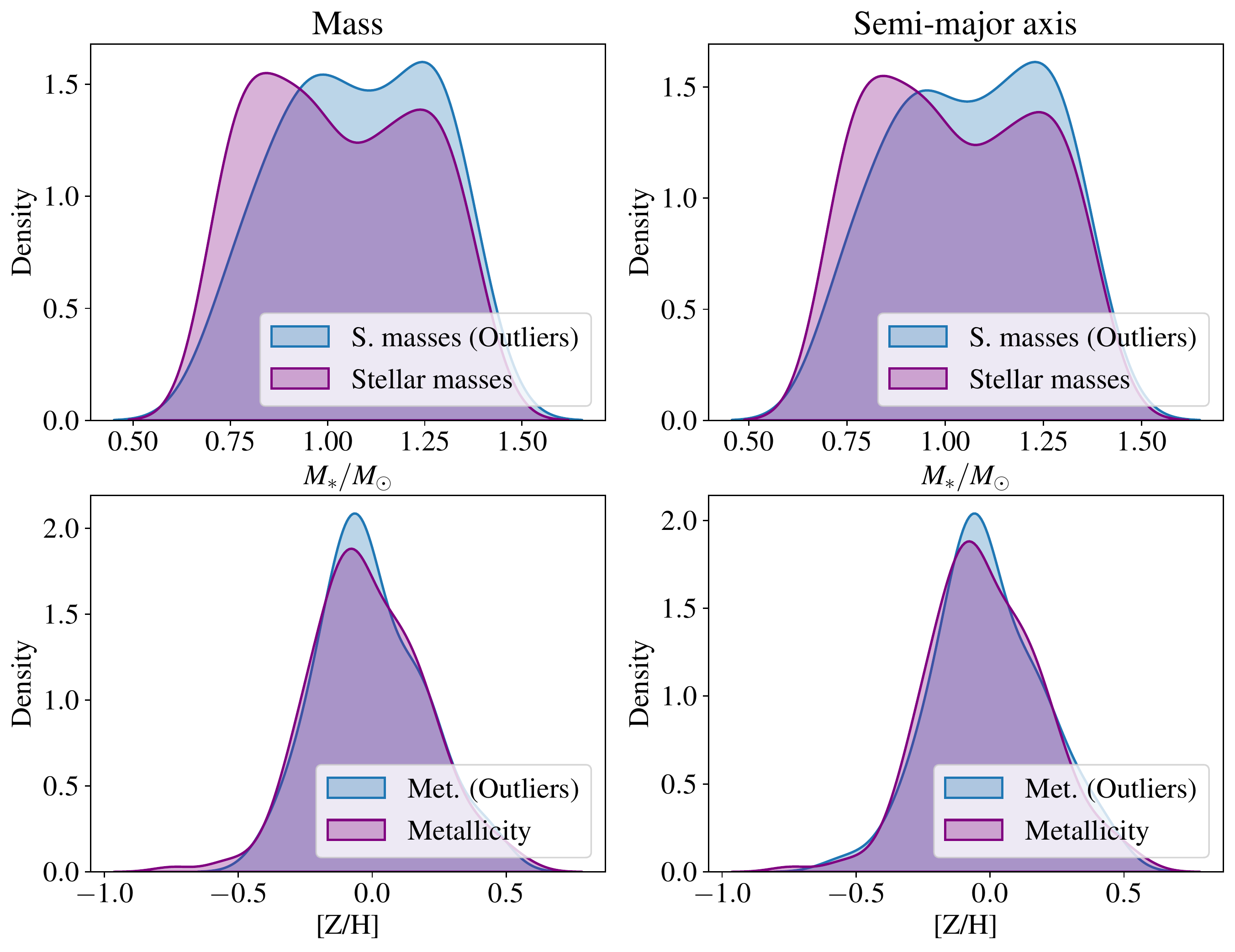}
    \caption{Stellar mass and metallicity distributions of the mass and semi-major axis outlier systems in the sim group: $10^6$ years.}
    \label{fig:st_out_1e6}
\end{figure*}

Figures~\ref{fig:st_out_1e5} and \ref{fig:st_out_1e6} show a comparison between stellar masses and metallicities of outlier systems and the whole sub-sample of systems evolving to $10^5$ and $10^6$ yr, respectively. From this comparison we can see that systems with higher stellar masses are more likely to change their orbital configuration in terms of planet mass and semi-major axis. This is likely due to the host star having a more significant gravitational influence over the orbital configuration of planets, whose initial locations are set by the local protoplanetary disk conditions and the result of migration processes and not from any secular relaxation. 

\begingroup
\renewcommand{\arraystretch}{1.5}
\begin{table}
	\centering
	\caption{Range of stellar masses and metallicities in which planetary mass outliers and semi-major axis outliers were found. The number of outliers compared to the total number of systems within the same stellar mass and metallicity range is reported below each value.}
	\subcaption*{Simulation group: $10^5$ years}
	\begin{tabular}{ccc}
	
	    \hline
	    \multirow{2}{*}{Mass} & $M_{\odot}=1.27^{+0.05}_{-0.05}$ & $\mathrm{[Z/H]}=0.27^{+0.11}_{-0.12}$ \\ & 5/23 systems & 5/29 systems \\
	    \hline
	    \multirow{2}{*}{Semi-major axis} & $M_{\odot}=1.14^{+0.18}_{-0.29}$ & $\mathrm{[Z/H]}=0.15^{+0.21}_{-0.16}$ \\ & 18/124 systems & 18/70 systems \\
	    \hline
	    
	\end{tabular}
	
	\bigskip
	
	\subcaption*{Simulation group: $10^6$ years}
	\begin{tabular}{ccc}
	    
	    \hline
	    \multirow{2}{*}{Mass} & $M_{\odot}=1.08^{+0.19}_{-0.19}$ & $\mathrm{[Z/H]}=0.03^{+0.27}_{-0.14}$ \\ & 22/97 systems & 22/117 systems \\
	    \hline
	    \multirow{2}{*}{Semi-major axis} & $M_{\odot}=1.07^{+0.26}_{-0.17}$ & $\mathrm{[Z/H]}=-0.06^{+0.37}_{-0.11}$ \\ & 11/112 systems & 11/136 systems \\
	    \hline

	\end{tabular}
	\label{tab:quantile_table}
\end{table}
\endgroup

In particular, for systems evolving to $10^5$ yr, significant planetary mass changes occurred in 22\% of the systems with stellar masses in the range $1.22 M_\odot < M_\star < 1.32 M_\odot$ and 17\% of the systems with metallicities between $0.15$ and $0.38$. Changes in semi-major axis occured for 15\% of the systems with stellar masses in the range $0.85 M_\odot < M_\star < 1.32 M_\odot$ and 26\% of the systems with metallicities between $-0.01$ and $0.36$.

Similarly, for systems evolving to $10^6$ yr, planetary mass changes happened for 23\% of the systems with with stellar masses in the range $0.89 M_\odot < M_\star < 1.27 M_\odot$ and 19\% of the systems with metallicities between $-0.11$ and $0.3$. Changes in semi-major axis occurred for 10\% of the systems with stellar masses in the range $0.9 M_\odot < M_\star < 1.33 M_\odot$ and for 8\% of the systems with metallicities between $-0.17$ and $0.31$. This is summarized in Table \ref{tab:quantile_table}. 

Similarly, the $10^5$ yr sub-sample shows that systems whose stars are metal-rich are more likely to change their orbital configuration in terms of mass and semi-major axis. This is likely due to the fact that metal-rich disks form more massive cores which in turn leads to a higher rate of giant planet formation. Giant planets are expected to be uncommon according to most planetary synthesis simulations, but they can significantly impact the orbital configuration of the planetary system which they inhabit.

This correlation of outliers with metallicity for our sub-sample evolving to 1 Myr seems to hold inasmuch outlier systems tend to have metallicities slightly higher than Solar. However, this is not as clear-cut as in the previous case ($10^5$ yr) since the original sub-sample is also slightly skewed toward higher metallicities. 

To summarize, the range of stellar parameters for outliers (Table \ref{tab:quantile_table}) show a narrow, skewed range relative to the full range of the simulations shown in Table \ref{Tab:initial_conditions}, further evidenced in the comparison between distributions in Figures \ref{fig:st_out_1e5} and \ref{fig:st_out_1e6}. This means that outliers do not appear as statistical flukes, but rather seem to come from an underlying distribution which may help evince a plausible physical explanation for orbital alteration.

We should note that as per our population synthesis simulations only 1 in 5 systems have giant planets, and therefore most of the dynamical simulations are performed with systems with lighter planets where orbital disruption is less likely. Further simulations with a larger sample of systems may be able to clear this up, but this lies outside of the exploratory scope of this work.

\subsection{Host stars of outliers vs. observations}\label{data-simulations_comparison}
 
Although our results above show that secular gravitational interactions seldom alter the final orbital configuration of a planetary system, outlier systems (i.e., systems that do change configuration) share enough commonalities that an inspection of their stellar parameters can inform us of the  conditions in which an orbital configuration change is more likely. In this section we explore the plausibility of observed systems having had a late, dynamic orbital evolution, by comparing the stellar parameters of outlier systems and those of the whole \textit{exoplanets.eu} sample of observed exoplanetary systems. 

\begin{figure*}
	\centering
	\begin{subfigure}{\textwidth}
		\centering
		\includegraphics[width=0.6\textwidth]{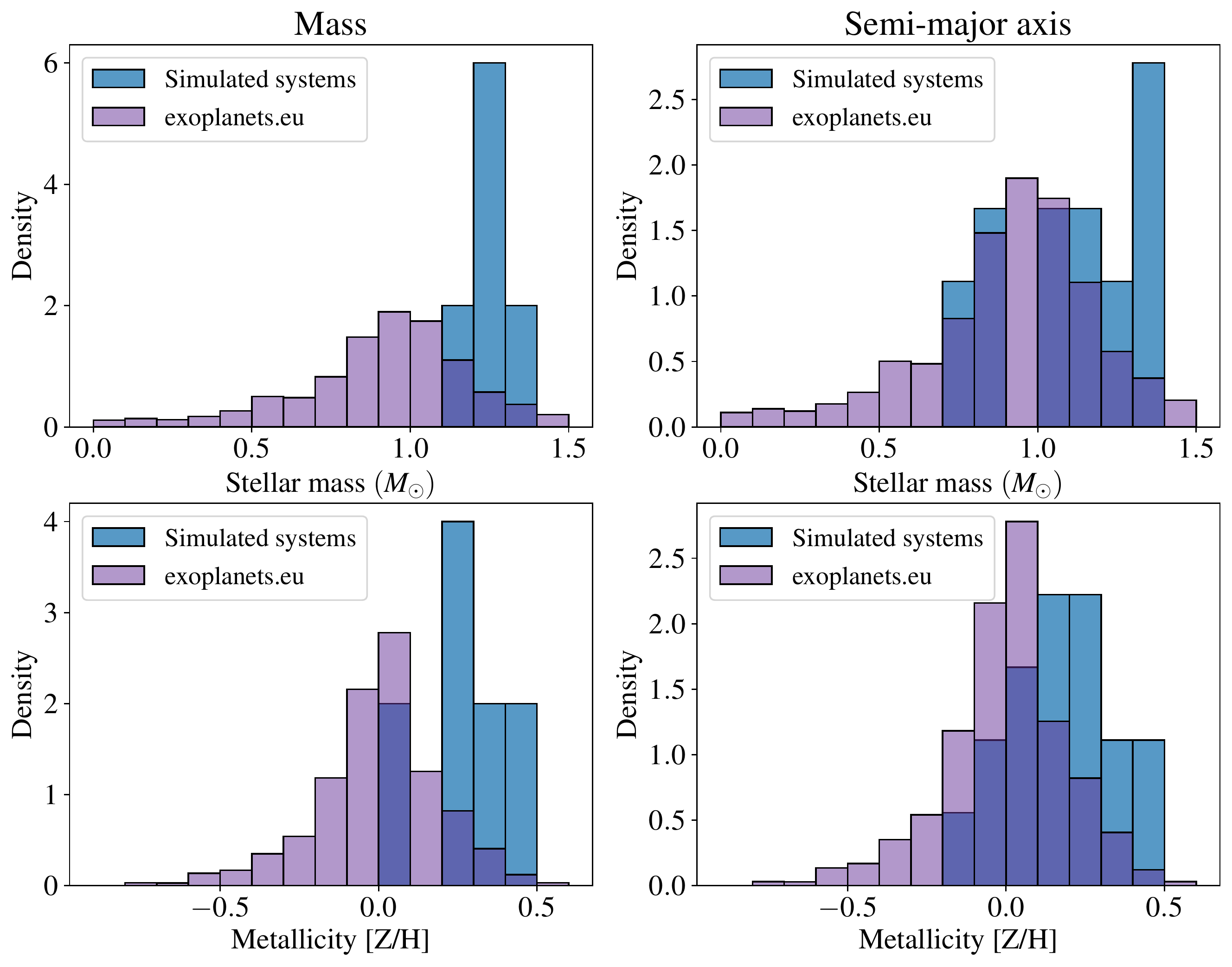}
		\caption{Systems evolving to $10^5$ yr.}
		\label{fig:star_out_hist_180}
	\end{subfigure}
	
	\begin{subfigure}{\textwidth}
		\centering
		\includegraphics[width=0.6\textwidth]{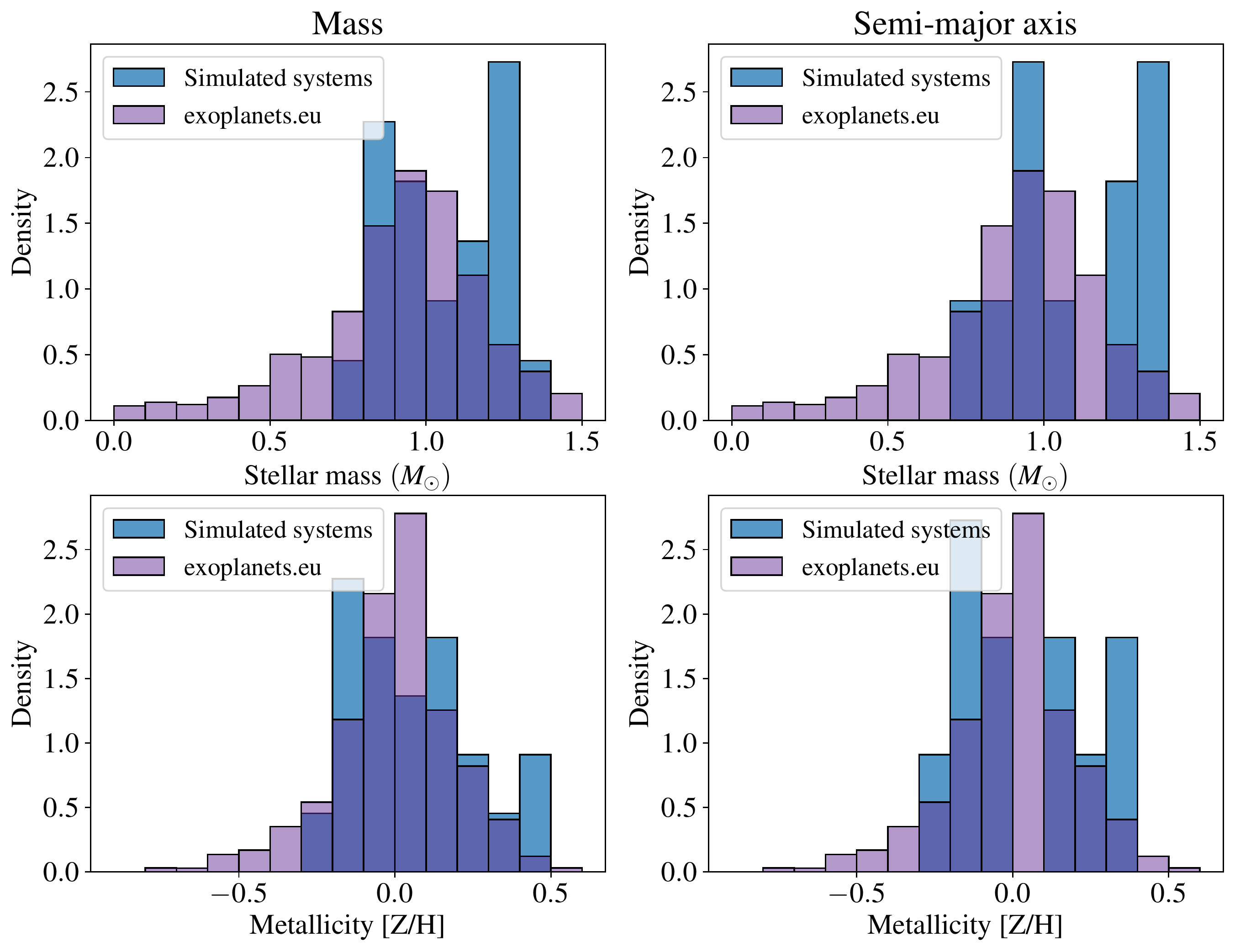}
		\caption{Systems evolving to $10^6$ yr.}
		\label{fig:star_out_hist_192}
	\end{subfigure}
	\caption{Stellar mass and metallicity distribution of outlier systems in the \textit{exoplanets.eu} database and simulated systems.}
	\label{fig:star_out_hist}
\end{figure*}

Figure \ref{fig:star_out_hist} shows a comparison of the normalized mass distribution and stellar metallicity of \textit{exoplanets.eu} systems vs. outliers. We make this direct comparison between differently biased system populations to point out that the range of stellar parameters where outliers may occur falls within the range of stellar parameters of observed systems, therefore showing that outliers do not form in unphysical or otherwise extremely pathological conditions.

Notice that the histogram density is proportional to the number of synthetic systems whose orbital configuration changed for a given range of stellar mass or metallicity. It is evident that outlier systems have stellar parameters within the range of observations, showing that it is plausible (although still unlikely) that some observed planetary systems have experienced significant orbital changes. In particular, it is unlikely that observed systems with host stars of spectral types K-M experienced a significant orbital change; a change in orbital configuration is more likely in systems around G- and F-type stars. 

Similarly, no outlier systems are found for metallicities much lower than solar: outliers can be found for host stars with metallicities above $-0.2$ with the probability peaking at $0.1-0.2$. Therefore, massive stars composed of more heavy elements (compared to Solar) are more likely to have planetary systems that have experienced substantial orbital changes during the dynamical evolution of planetary systems.

We should emphasize that differences in the observed and simulated populations as illustrated here may be due to observational biases affecting the distribution of the observed population, and thus observations of massive and metal-rich stars may not necessarily be more likely to reveal signs of center-of-mass changes at the late stages of planet formation identified by our simulations.

\subsection{Planetary system reclassification after gravitational evolution}

We end this discussion of results with an analysis inspired by our center-of-mass classification scheme from Section~\ref{Class}. In our method described above, we use the posterior distributions from the GMM clustering scheme in order to check whether a system is reclassified after considering the center-of-mass uncertainty. Here we borrow the basics from this analysis in order to check whether a system is reclassified after its secular evolution. We should clarify that reclassification due to dynamical evolution and reclassification due to measurement errors cannot be directly compared, as the former refers to the same synthetic system evolving and possibly changing during a time span of more than $10^5$ yr and the latter refers to whether we know enough about an observed system to properly classify it in terms of its $r_\text{cm}$.

As expected from our results (summarized in the previous section), for most cases the center of mass $r_\text{cm}$ of the evolved system is still close to its original value and therefore they are not reclassified. However, for outlier systems the semi-major axis and planetary mass distributions change and therefore their center of mass also changes, leading to a reclassification. We shall take a look at the results from this analysis, with \emph{planet mass outliers} meaning systems whose planet mass configuration changed, and \emph{semi-major axis outliers} meaning systems that experienced changes in their semi-major axis configurations after secular evolution.

\begin{itemize}
    \item \emph{Outliers in systems evolving to $10^5$ yr.} $33 \% $ of the semi-major axis outliers and 20\% of the planet mass outliers are reclassified to \emph{Sub-Mercurian} from other clusters (see the original clustering of synthetic systems in Figure \ref{fig:classif_sy}). Thus, outliers produce a down-shift in the center of mass $r_\text{cm}$ towards the  $0.2-1$ AU region. We also noticed that planet mass outliers often occur in systems that lose planets in gravitational interactions with massive planets.
    \item \emph{Outliers in systems evolving to $10^6$ yr.}  $54 \%$ of semi-major axis outliers are reclassified, but not to any particular cluster. $40\%$ of planet mass outliers are reclassified, most of them as \emph{Sub-Mercurian}. To wit, when there is a loss of a planet (or more), the center of mass decreases, whereas a shift in the semi-major axis configuration may not necessarily lead to planet loss, so the center of mass shift can be in any direction.
\end{itemize}

\section{Conclusions}\label{conclusion}

In this work we are interested in finding the conditions which set apart planetary systems that significantly change their planet mass and semi-major axis configurations after allowing for dynamical evolution. To this end, we evolve $\sim400$ synthetic planetary systems with $\sim5000$ planets forward in time using an adaptive-step collisional N-body simulator. Our seed planetary systems were created from a Monte Carlo planet population synthesis model, using the star-disk distributions in Table~\ref{Tab:initial_conditions} as initial conditions.

In order to better understand the architecture of observed and synthetic planetary systems and how it may change we first built a clustering scheme for the center of mass $r_\text{cm}$ (as defined in Section~\ref{Class}) based on Gaussian Mixture Models and Information Criteria. In the case of observed planetary systems we included an additional model validation step based on Approximate Bayesian Computation methods.

Our methods yielded four groups of observed planetary systems: \textit{Sub-Mercurian} for $10^{-2}$ AU $\leq r_\text{cm}< 10^{-1}$ AU with $46.0\%$ of observed systems, \textit{Venusian} for $10^{-1}$ AU $\leq r_\text{cm} < 1$ AU with $12.8\%$ of observed systems, \textit{Solar-like} for $1\ \text{AU} \leq r_\text{cm} < 10$ AU with $37.0\%$ of observed systems, and \textit{Peripheral} for $10$ AU $\leq r_\text{cm}$ with $4.20\%$ of observed systems. Our Approximate Bayesian Computation validation method found that our scheme is robust, with less that 2\% of systems being re-classified when taking measurement errors into account.

We found that synthetic planetary systems are also best classified by 4 clusters of $r_\text{cm}$,  corresponding to the same clusters found for the observed systems (within a factor 2), albeit with radically different posterior amplitudes that are due to observational biases. 

In order to assess the possible change of orbital configuration in a system after gravitational evolution we used a statistical framework based on a cross-validated Kernel Density Estimation of the planet mass, semi-major axis, and eccentricity and a calculation of the Kullback-Leibler divergence between the before-and-after distributions as a Bayesian measurement of the statistical distance between the initial and final orbital configurations.

After letting the synthetic systems evolve gravitationally to $10^5$ and $10^6$ yr we found that $\gtrsim 90 \%$ of the systems present a stable dynamical evolution, that is, their orbital parameters (planet mass, eccentricity, semi-major axis) do not significantly change during the latter stages of planetary formation.

However a few systems ($\lesssim 10 \%$) had statistically significant changes, which we refer to in the text as \emph{outliers}: planetary systems whose KL divergence is above $1\sigma$ above the mean. Further analysis of these outliers shows that even though they may appear at almost any realistic stellar mass or metallicity, systems with F- and G- type host stars are much more likely to experience changes in their orbital configuration when compared to systems with K- and M- type host stars. This is compatible with the Nice Model \cite{Tsiganis2005}, which suggests a scenario for the young Solar System in which its original orbital configuration changed due to gravitational interactions with leftover planetesimals. Likewise, systems with metallicities higher than solar ([Fe/H] $> 0.2$) are also more likely to change their orbital configuration due to interactions with massive planets. This is mostly due to the fact that giant planets are more likely to be found in systems with high stellar metallicity.

We noticed differences between the distribution of outliers vs. the rest of synthetic systems and vs. observed systems. On one hand, the former is significant because it tells us that the specific mass and metallicity conditions shown here favor orbital configuration changes. On the other hand, the latter may be due to observational biases, which means that further observations of already discovered massive, metal-rich stars may not be as informative as discovering new planet-hosting stars.

Finally, we check whether outliers are reclassified in our clustering scheme after dynamical evolution. Interestingly, only about 20-50\% outliers (less than 10 systems out of nearly 400 total) are reclassified, meaning that statistically significant changes in $r_\text{cm}$ do not necessarily translate to significant changes in cluster membership. Our classification scheme is thus proven to be fairly robust with respect to observational uncertainties and gravitational evolution effects.

Since outliers are usually classified as Sub-Mercurian in terms of their center of mass, we conclude that orbital configuration changes are related to the presence of massive, migrated planets. These close-in giants alter the orbital stability of newly born planetary systems, in many cases causing scenarios where other planets are ejected from the system. 

In future work we will focus on running more simulations from our original pool of synthetic systems (and for longer timescales). Thus we will more thoroughly sweep the parameter space to better quantify the precise causes behind changes in orbital configurations brought about by secular gravitational interactions.

\section*{Acknowledgements}

The authors would like to thank: Yamila Miguel for providing the first version of the Planet Population Synthesis code, Jaime Forero-Romero for giving us access to the UniAndes computing facilities in which this code was run to produce our synthetic planetary system dataset, Pablo Cuartas and Mateo Restrepo for providing us with the first version of the N-body collisional simulation code, Santiago Vargas and Mario Acero for their valuable feedback. We would also like to thank the anonymous referee for their insight, which helped convey our conclusions more clearly.

% %%%%%%%%%%%%%%%%%%%%%%%%%%%%%%%%%%%%%%%%%%%%%%%%%%
\section*{Data Availability}

Initial data from the simulations can be found at: \url{https://github.com/saint-germain/population_synthesis/tree/master/data_from_sim}. The rest of the data and code are available upon request.



\bibliographystyle{mnras}
\bibliography{mnras_template} % if your bibtex file is called example.bib

%\begin{thebibliography}{99}
%\end{thebibliography}

%%%%%%%%%%%%%%%%%%%%%%%%%%%%%%%%%%%%%%%%%%%%%%%%%%

%%%%%%%%%%%%%%%%% APPENDICES %%%%%%%%%%%%%%%%%%%%%

% \appendix

% \section{Some extra material}

% If you want to present additional material which would interrupt the flow of the main paper,
% it can be placed in an Appendix which appears after the list of references.

%%%%%%%%%%%%%%%%%%%%%%%%%%%%%%%%%%%%%%%%%%%%%%%%%%

% Don't change these lines
\bsp	% typesetting comment
\label{lastpage}
\end{document}